\documentclass[twocolumns,
reprint,
superscriptaddress,
amsmath,
amssymb,
prx,
nofootinbib
]{revtex4-2}

\usepackage{
graphicx,
amsmath,
amsfonts,
amssymb,
amsthm,
mathtools,
xcolor, 
algpseudocode,
listings,
dsfont,
verbatim,
comment,
hyperref,
cancel,
subcaption,
physics,
thm-restate,
quantikz
}

\usepackage[ruled, vlined, linesnumbered, commentsnumbered]{algorithm2e}
\usepackage[capitalize]{cleveref}
\usepackage[font=small,labelfont=bf]{caption}
\newtheorem{theorem}{Theorem}[section]

\newtheorem{example}{Example}[section]
\newtheorem{lemma}{Lemma}[section]
\theoremstyle{remark}

\DeclareMathOperator*{\argmin}{arg\,min}

\begin{document}

\title{The quantum super-Krylov method}

\author{Adam Byrne}
\email{byrne.adam@ibm.com}
\affiliation{IBM Quantum, IBM Research Europe - Dublin, IBM Technology Campus, Dublin 15, Ireland}
\affiliation{School of Mathematics, Trinity College
Dublin, Ireland}

\author{William Kirby}
\affiliation{IBM Quantum, IBM T.J. Watson Research Center, Yorktown Heights, NY 10598, United States}

\author{Kirk M. Soodhalter}
\affiliation{School of Mathematics, Trinity College
Dublin, Ireland}

\author{Sergiy Zhuk}
\affiliation{IBM Quantum, IBM Research Europe - Dublin, IBM Technology Campus, Dublin 15, Ireland}

\begin{abstract}

The problem of estimating the ground-state energy of a quantum system is ubiquitous in chemistry and condensed matter physics. Krylov quantum diagonalization (KQD) has emerged as a promising approach for this task. However, many KQD methods rely on subroutines, particularly the Hadamard test, that are challenging to implement on near-term quantum computers. We present a novel KQD method that uses only real-time evolutions and recovery probabilities, making it well adapted for existing quantum hardware. The method entails numerical differentiation in post-processing, and so we present a novel derivative estimation algorithm that is robust to noisy data. Under assumptions on the spectrum of the Hamiltonian, we prove that our algorithm converges exponentially quickly to the ground-state energy and present a numerical demonstration using tensor network simulations.

\end{abstract}

\maketitle

\section{Introduction}

One of the major computational challenges in chemistry and physics is approximating the ground-state energy of many-body quantum systems. In general, this task is exponentially hard on classical computers, due to the cost of classically representing vectors in the system's Hilbert space \cite{simQuantumCompGuzik,hardwareEffKandala,faultTolerantYuan}. Many quantum algorithms avoid this storage problem by encoding the Hilbert space in the exponentially-sized Hilbert space of a register of qubits ~\cite{abelianStabiliserKitaev, VQEperuzzo, virtualSubspaceUrbanek, lanczosGreensFnBaker, epperly2021subspacediagonalization, davidsonTkachenko, groundEnergyEstDong, heisenGroundEstLin, adaptiveHaoya23, unifiedNoiseYizhi2023, ESPRIT2024, quantumFilterParrish, multireferenceStair, filterDiagJeffery, realTimeEvolKlymko, cortes2022krylov, quantumPowerSeki, realTimeKrylovShen, exactLanczosKirby, eigenstatesMotta, Motta_2024, diagonalisationOfManyBodyYoshioka, yu2025quantumcentricalgorithmsamplebasedkrylov, robledomoreno2024chemistryexactsolutionsquantumcentric}. One such approach, \textit{Quantum Subspace Diagonalization} (QSD) \cite{virtualSubspaceUrbanek, lanczosGreensFnBaker, epperly2021subspacediagonalization, davidsonTkachenko, groundEnergyEstDong, heisenGroundEstLin, adaptiveHaoya23, unifiedNoiseYizhi2023, ESPRIT2024,quantumFilterParrish,multireferenceStair,filterDiagJeffery,realTimeEvolKlymko,cortes2022krylov,quantumPowerSeki,realTimeKrylovShen,exactLanczosKirby,eigenstatesMotta,Motta_2024,diagonalisationOfManyBodyYoshioka,yu2025quantumcentricalgorithmsamplebasedkrylov,robledomoreno2024chemistryexactsolutionsquantumcentric}, has shown promise for implementation on near-term quantum devices. These methods project the system Hamiltonian into a low-dimensional subspace of the Hilbert space, and then obtain approximate energy levels within that subspace. Variants of these methods have been experimentally demonstrated for the 56-qubit Heisenberg model \cite{diagonalisationOfManyBodyYoshioka}, and the 85-qubit Anderson model \cite{yu2025quantumcentricalgorithmsamplebasedkrylov}.

We outline the general QSD procedure. Let $H$ be the Hamiltonian for an $n$-qubit many-body system, with eigenvalue equation
\begin{equation}
    H \ket{u} = \lambda \ket{u}
\end{equation}
where $\lambda_0\leq...\leq\lambda_{N-1}$, for $N=2^n$. Denote by $\mathcal{H}$ the $N$-dimensional Hilbert space of the system, and let $\mathcal{A}$ be an $m$-dimensional subspace of $\mathcal{H}$ with basis vectors $A=\left\{\ket{\phi_0},...,\ket{\phi_{m-1}}\right\}$. QSD obtains approximate eigenpairs $(\widetilde{\lambda},\widetilde{u})$ of $H$ via the Galerkin projection:
\begin{equation}\label{eq:H-galerkin-proj}
    \boxed{\text{Find $\widetilde{\lambda} \in \mathbb{C}$ and $\widetilde{u} \in \mathcal{A}$ s.t. $H\widetilde{u} - \widetilde{\lambda} \widetilde{u}\perp\mathcal{A}$.}}
\end{equation}
It can be shown that the quantities $\widetilde{\lambda}$ in \eqref{eq:H-galerkin-proj} solve the generalized eigenvalue problem
\begin{equation}\label{eq:projection of original hamiltonian}
    \mathbf{H}c = \widetilde{\lambda} \, \mathbf{S} c
\end{equation}
where $\mathbf{H}$ and $\mathbf{S}$ are order-$m$ Toeplitz-Hermitian matrices with entries
\begin{equation}\label{eq:entries of Hproj S}
\begin{split}
    \mathbf{H}_{jk} = \bra{\phi_j} H \ket{\phi_k}, 
    \quad \mathbf{S}_{jk} = \bra{\phi_j} \ket{\phi_k}.
\end{split}
\end{equation}
In practice, $\mathbf{H}$ and $\mathbf{S}$ are constructed entry-wise on a quantum computer, and the problem \eqref{eq:projection of original hamiltonian} is solved using classical methods (e.g., the QZ algorithm \cite{golub2013matrix}). The lowest eigenvalue $\widetilde{\lambda}_0$ of \eqref{eq:projection of original hamiltonian} then serves as an estimate of the true ground-state energy.

The choice of subspace is key to the algorithmic and convergence properties of QSD. The standard choice in classical computing is the Krylov subspace basis ${A=\{\ket{\phi_0}, H\ket{\phi_0},....,H^{m-1}\ket{\phi_0}\}}$. The resulting diagonalization algorithm is referred to as the \textit{Lanczos} method, and has been shown to achieve exponential convergence with respect to the subspace dimension \cite{saadConvergence1980}. However, the operation $\ket{\phi_0} \mapsto H^k \ket{\phi_0}$ is non-unitary in general, and so cannot be directly implemented by quantum circuits. Many alternative ``quantum-friendly'' subspaces have been proposed in recent years. For example, the \textit{QLanczos} method \cite{eigenstatesMotta} uses the Krylov subspace generated by powers of the imaginary time evolution operator $e^{Ht}$. The operation $\ket{\phi_0}\mapsto e^{Ht}\ket{\phi_0}$ is non-unitary, but can be approximated on near-term devices using the quantum imaginary time evolution (QITE) algorithm. However, QITE is limited by the unfavorable growth in circuit depth with respect to the number of timesteps \cite{Motta_2024}. Another subspace choice, proposed in \cite{exactLanczosKirby}, is that generated by the Chebyshev polynomials of $H$. The subspace generated by this approach is exactly equivalent to the subspace used in the Lanczos method, up to sampling noise. However, its practical implementation involves block-encoding unitaries, which require ancilla qubits and deep quantum circuits, and so the approach is more suited to fault-tolerant quantum devices \cite{Motta_2024}.

In the present work, we focus on subspaces generated by powers of the real-time evolution operator ${U(t)\coloneq e^{-\imath H t}}$, that is,
\begin{equation}\label{eq:krylov space for H}
    A =\{\ket{\phi_0},U(t)\ket{\phi_0},...,U(t)^{m-1}\ket{\phi_0}\},
\end{equation}
for some fixed $t\in\mathbb{R}$. The resulting diagonalization method is referred to in the literature as \textit{Krylov Quantum Diagonalization} (KQD). The advantage of this approach is that, for sufficiently simple and structured Hamiltonians, the unitary operation $\ket{\phi_0}\mapsto U(t)\ket{\phi_0}$ can be efficiently approximated by circuits that are shallow enough to be performed on near-term devices. Furthermore, it can be shown that, under ideal conditions, the projection error in the associated estimate $\widetilde{\lambda}_0$ converges exponentially quickly with the subspace dimension $m$, even in the presence of noise \cite{epperly2021subspacediagonalization}.

However, there is a catch when it comes to practical implementation. Using the subspace in \eqref{eq:krylov space for H}, the projected matrices become
\begin{equation}\label{eq:entries of Hproj S real-time}
\begin{split}
    \mathbf{H}_{jk}(t) &= \bra{\phi_0} U^{-j}(t) H U^{k}(t) \ket{\phi_0}, \\
    \mathbf{S}_{jk}(t) &= \bra{\phi_0} U^{-j}(t) U^{k}(t) \ket{\phi_0}.
\end{split}
\end{equation}
The standard method for computing complex overlaps
of the form \eqref{eq:entries of Hproj S real-time} on a quantum device is the Hadamard
test \cite{Nielsen_Chuang_2010}. However, in general, this procedure involves controlled unitary operations, which require circuits that are prohibitively deep for near-term quantum computers \cite{nisqPreskill,nisqLau,overlapNisqMitarai}. Recent work has shown that the Hadamard test can be avoided for Hamiltonians with particular structure. For instance, if the Hamiltonian has particle number symmetry, or total spin or spin projection symmetries, then hybrid quantum-classical multi-fidelity protocols \cite{cortes2022krylov} can be used to efficiently estimate the entries in \eqref{eq:entries of Hproj S real-time}. Furthermore, for systems with sparse ground states, sampling-based KQD methods \cite{yu2025quantumcentricalgorithmsamplebasedkrylov} can obtain approximate energies using only sampling in the computational basis.

We present a novel addition to this class of ``near-term friendly" KQD methods. Our approach, which we call the \textit{super-Krylov} method, requires only real-time evolutions, recovery probabilities, and a classical derivative estimation algorithm in post-processing. This approach can be applied to two classes of Hamiltonians; where $\{H,W\}=0$ for some unitary (or anti-unitary) $W$, or where the most-excited energy of $H$ is known. The method can also be applied to estimate the spectral range of a general Hamiltonian, provided $\ket{\phi_0}\mapsto e^{-\imath H t} \ket{\phi_0}$ admits efficient implementation via quantum circuits. We provide rigorous upper bounds on the projection error on the super-Krylov energy estimate, and show that this error converges exponentially quickly with the subspace dimension.

For classical post-processing, we present a novel derivative estimation algorithm that is robust to noisy data. This approach does not assume \textit{a priori} knowledge of the statistical distribution of the noise, and so can be applied irrespective of the noise description of the quantum device. The present differentiation algorithm can be applied in the general setting, given smoothness assumptions on the input signal. We derive the derivative estimate via a regularized minimization problem and show that it admits a closed-form analytical expression. Finally, we provide a proof of convergence and establish a ``worst-case" upper bound on the estimation error.

The paper is structured as follows. In \cref{sec:super-krylov-method}, we present the super-Krylov method and provide an algorithmic framework for its practical implementation. In \cref{sec:derivative-estimation}, we present a novel derivative estimation method for use in post-processing. We also derive an analytical form for the derivative estimate and provide a proof of convergence. In \cref{sec:error-analysis}, we present an analysis of the projection error associated with the super-Krylov method, and in \cref{sec:experiments}, we demonstrate the method numerically using tensor networks.

\section{The super-Krylov method}\label{sec:super-krylov-method}

The super-Krylov method approximates the eigenvalues of the \textit{Liouvillian super-operator} $\mathfrak{L}:\mathcal{M}_N \to \mathcal{M}_N$, defined by
\begin{equation}
\begin{split}
    \mathfrak{L}X &\coloneqq [H,X] \\
    &= HX - XH,
\end{split}
\end{equation}
where $\mathcal{M}_N$ is the Hilbert space of $N\times N$ complex matrices equipped with the trace inner product. The time evolution of a density matrix $\rho(t)$ in a closed system with Hamiltonian $H$ is governed by the \textit{Liouville-von Neumann equation}:
\begin{equation}\label{eq:liouville-von-neumann}
    \imath \frac{d}{dt}\rho(t) = \mathfrak{L}\rho(t), \quad \rho_{in}\coloneqq \rho(0).
\end{equation}
For a review of the Liouville formalism of quantum mechanics, see \cite{Gyamfi_2020}. The super-operator $\mathfrak{L}$ has eigenvalue equation
\begin{equation}\label{eq:def-J}
    \mathfrak{L} \ketbra{u_p}{u_q} = (\lambda_ p - \lambda_q) \ketbra{u_p}{u_q},
\end{equation}
for $p,q=0,...,N-1$, with smallest eigenvalue ${\Delta_0 \coloneqq (\lambda_0 - \lambda_{N-1})}$. 

Let $\mathcal{B}$ be an $m$-dimensional space in $\mathcal{M}_N$, with basis states $B=\{\rho_0, ..., \rho_{m-1}\}$. Approximate eigenpairs of $\mathfrak{L}$ are obtained via the Galerkin projection:
\begin{equation}\label{eq:J-galerkin-proj}
    \boxed{\text{Find $\widetilde{\Delta}\in\mathbb{C}$ and $\widetilde{X} \in \mathcal{B}$ s.t. $\mathfrak{L}\widetilde{X} - \widetilde{\Delta}\widetilde{X} \perp \mathcal{B}$.}}
\end{equation}
Similarly to standard KQD, the quantities $\widetilde{\Delta}$ in \eqref{eq:J-galerkin-proj} solve the generalized eigenvalue problem
\begin{align}\label{eq:J projected eigenvalue problem}
    \mathbf{J}a  = \widetilde{\Delta} \, \mathbf{R}a
\end{align}
where $\widetilde{\Delta}_0\leq...\leq\widetilde{\Delta}_{m-1}$ and $\mathbf{J}$ and $\mathbf{R}$ are order-$m$ matrices with entries
\begin{equation}\label{eq:J R}
    \begin{split}
        \mathbf{J}_{jk} = \text{Tr} ([H,\rho_j]^\dagger \rho_k), \quad 
        \mathbf{R}_{jk} = \text{Tr}(\rho_j^\dag \rho_k ).
    \end{split}
\end{equation}
The derivation of \eqref{eq:J R} is deferred to \cref{sec-app:super-k}. We have just to choose the basis states $B$. Motivated by the subspace choice in standard KQD, we take the basis states generated by powers of the real-time evolution operator, that is
\begin{equation}\label{eq:density matrix basis}
    \rho_j(t) = U^j(t) \rho \, U^{-j}(t), \quad j=0,...,m-1,
\end{equation}
for some initial state $\rho=\ketbra{v}{v}$. With the choice, the projected problem in \eqref{eq:J projected eigenvalue problem} becomes
\begin{equation}\label{eq:Jt-Rt-proj-prob}
    \mathbf{J}(t) a = \widetilde{\Delta}\mathbf{R}(t) a
\end{equation}
where
\begin{align}
        \mathbf{J}_{jk}(t) &= \text{Tr} \big([H,\rho_j(t)] \, \rho_k(t) \big) \label{eq:J(t)}, \\ 
        \mathbf{R}_{jk}(t) &= \text{Tr}\big(\rho_j(t) \rho_k(t) \big) \label{eq:R(t)}.
\end{align}
By the cyclic property of the trace, \eqref{eq:R(t)} can be written
\begin{equation}\label{eq:R-probabilities}
    \mathbf{R}_{jk}(t) = |\langle v | U^{-j}(t) U^k(t) | v \rangle |^2.
\end{equation}
Similarly, \eqref{eq:J(t)} can be expanded as
\begin{equation}\label{eq:J-toeplitz}
    \mathbf{J}_{jk}(t) = 2 \imath  \mathfrak{Im} \big( \bra{v}U^{-k}(t) U^j(t)\ketbra{v}{v}U^{-j}(t) H U^k(t) \ket{v} \big).
\end{equation}
Note that $\mathbf{R}_{jk}(t) \in \mathbb{R}$ and $\mathbf{J}_{jk}(t) \in \imath \mathbb{R}$. If the time evolution is performed exactly, then ${[H,U(t)] = 0}$, and the matrices $\mathbf{J}(t)$ and $\mathbf{R}(t)$ are both Toeplitz-Hermitian. 

Assume $\ket{v}=Q\ket{0}$, for some unitary $Q$ that can be efficiently implemented using quantum circuits. Then, the quantities in \eqref{eq:R-probabilities} can be efficiently estimated by performing $\ket{0}\mapsto Q^\dag U^{k-j}(t) Q \ket{0}$, and then sampling the probability of observing $|00...0\rangle$ in the resulting state. The corresponding quantum circuit is presented in \cref{fig:R-circuit}. It remains to show how one may efficiently estimate the entries in \eqref{eq:J-toeplitz}. Using \eqref{eq:liouville-von-neumann}, one readily obtains
\begin{equation}\label{eq:derivative-relation}
    \mathbf{J}_{jk}(t) = \frac{\imath}{(k-j)} \frac{d}{d t} \mathbf{R}_{jk}(t).
\end{equation}
Thus, the entries of $\mathbf{J}(t)$ can be obtained directly from the time derivative of the entries of $\mathbf{R}(t)$. The derivation of \eqref{eq:derivative-relation} is deferred to \cref{sec-app:super-k}.

\cref{alg:super-Krylov} presents the computational protocol for estimating the entries of $\mathbf{J}(t)$ and $\mathbf{R}(t)$, given the representation in \eqref{eq:R-probabilities} and \eqref{eq:derivative-relation}. In practice, the time parameter is taken as some fixed value $t=\mathrm{d}t>0$ and we solve the resulting problem $(\mathbf{J},\mathbf{R})\equiv\bigl(\mathbf{J}(\mathrm{d}t),\mathbf{R}(\mathrm{d}t)\bigr)$. Details on the choice of time step $\mathrm{d}t$ are discussed in \cref{sec:error-analysis}.

In step 3 of \cref{alg:super-Krylov}, the Toeplitz-Hermitian structure of $\mathbf{J}$ and $\mathbf{R}$ is used to construct each matrix from their first row. However, in practice, the time evolution $\ket{v}\to U(t)\ket{v}$ is approximated using some discretization technique (e.g., the Trotter-Suzuki decomposition \cite{Hatano_2005}). Furthermore, the sampling procedure for measuring $z$ in step 1 will incur Monte Carlo noise \cite{Nielsen_Chuang_2010}. As a result, the matrices $\mathbf{J}$ and $\mathbf{R}$ will not be Toeplitz-Hermitian, so the approximation $(\widehat{\mathbf{J}},\widehat{\mathbf{R}})$ obtained by enforcing this structure will represent the true matrices corrupted by discretization and sampling errors \cite{epperly2021subspacediagonalization}.

In step 5 of \cref{alg:super-Krylov}, ``eigenvalue thresholding'' is employed to regularize the approximate Krylov matrix pair $\big(\widehat{\mathbf{J}},\widehat{\mathbf{R}}\big)$ before it is solved \cite{epperly2021subspacediagonalization}. This is required because $\mathbf{R}$ is ill-conditioned, both in general and in practice, and so the noisy matrix $\widehat{\mathbf{R}}$ can be singular~\cite{epperly2021subspacediagonalization,kirby2024analysis}. Eigenvalue thresholding avoids ill-conditioning of the problem $\big(\widehat{\mathbf{J}},\widehat{\mathbf{R}}\big)$ by projecting out the eigenspaces of the Gram matrix $\widehat{\mathbf{R}}$ whose eigenvalues are below some threshold $\epsilon>0$, thus enforcing that the $\epsilon$-thresholded matrix $\widehat{\mathbf{R}}$ is positive definite.

\begin{algorithm}
    \caption{The super-Krylov method.}
    \label{alg:super-Krylov}
    \KwData{Hamiltonian ${H}$, initial state $\ket{v}$, Krylov dimension $m$, timestep $\mathrm{d}t$, number of datapoints $D$, spectral threshold $\epsilon>0$.} 
    \KwResult{Estimate $\widehat{\Delta}_0$ of the minimum eigenvalue of $\big(\mathbf{J}(\mathrm{d}t), \mathbf{R}(\mathrm{d}t)\big)$.}
    \BlankLine
    \BlankLine
    Measure $z(t)\coloneq \left|\bra{v} U(t) \ket{v} \right|^2$ at $t\in\mathcal{T}\coloneqq\{0=t_1<t_2<...<t_D=m \,\mathrm{d}t\}$\;
    Construct \textit{classical} continuous function estimates of $z(t)$ and $\frac{d}{dt}z(t)$ over $t\in[0,m\mathrm{d}t]$, denoted by $\widehat{z}(t)$ and $\widehat{\frac{d}{dt}z}(t)$ respectively\;
    Construct an estimate $(\widehat{\mathbf{J}},\widehat{\mathbf{R}})$ of $({\mathbf{J}},\mathbf{R})$: Take $\widehat{\mathbf{R}}_{00}\leftarrow 1$, $\widehat{\mathbf{J}}_{00}\leftarrow0$, and classically estimate the first rows via 
    \begin{align}
        \widehat{\mathbf{R}}_{0k}(t) &\leftarrow \left.\widehat{z}(t)\right|_{t=k\mathrm{d}t}, \\ 
        \widehat{\mathbf{J}}_{0k}(t) &\leftarrow \left.\frac{\imath}{k}\widehat{\frac{d}{dt}z}(t)\right|_{t=k\mathrm{d}t},
    \end{align}
    for $k=1,...,m-1$. Impute the remaining entries of $\widehat{\mathbf{J}}$ and $\widehat{\mathbf{R}}$ by enforcing the Toeplitz-Hermitian structure\;
    Perform eigenvalue thresholding to obtain the $\epsilon$-thresholded pair $(\widehat{\mathbf{J}}_\epsilon,\widehat{\mathbf{R}}_\epsilon)$\;
    Solve the generalized eigenvalue problem
    \begin{equation}
        \left\{(\widehat{\Delta}, a)\right\}\leftarrow\textup{Eig}\left(\widehat{\mathbf{J}}_\epsilon,\widehat{\mathbf{R}}_\epsilon\right)
        ;
    \end{equation}
    \\
    $\widehat{\Delta}_0 \leftarrow \min_k \widehat{\Delta}_k$\;
\end{algorithm}

\begin{figure}
    \centering
    \begin{quantikz}[row sep={0.8cm,between origins}, column sep=0.7cm]
    \lstick{$\ket{0}^{\otimes n}$} & \gate{Q} 
                                    & \gate{U(t_k)} 
                                    & \gate{Q^\dagger} 
                                    & \meter{} 
                                    \\
    \end{quantikz}
    \caption{Circuit for sampling $z(t)=|\bra{v}U(t)\ket{v}|^2$ at the timepoints $t=t_k$ in \cref{alg:super-Krylov}.}
    \label{fig:R-circuit}
\end{figure}

\subsection{Applications}\label{subsec:superk-applications}

We have shown that the super-Krylov method enables the efficient estimation of $\Delta_0 = \lambda_0 - \lambda_{N-1}$, which in turn admits the following applications:
\begin{enumerate}
    \item Estimation of the spectral range $|\Delta_0|$ of $H$.
    \item Estimation of the ground-state energy of two classes of Hamiltonians $H$, where either:
    \begin{enumerate}
        \item there exists a unitary (or anti-unitary) operator $W$ such that $\{H,W\} = 0$.
        \item the most excited energy $\lambda_{N-1}$ of $H$ is known.
    \end{enumerate}
\end{enumerate}
If condition $2.\, (a)$ holds, then $H$ has a symmetric spectrum, and so $\lambda_0 = \Delta_0/2$. If condition $2.\,(b)$ holds, then the ground-state energy may be estimated via ${\lambda_0 = \Delta_0 + \lambda_{N-1}}$. The estimation of the spectral range finds application in many quantum algorithm subroutines. For example, one may constrain the spectrum of $H$ to lie within the interval $[-1,1]$ by performing the transformation $H \mapsto H / |\Delta_0|$. This transformation is performed in qubitization \cite{low2019hamiltonian}, quantum phase estimation \cite{abelianStabiliserKitaev}, and quantum linear system solvers \cite{PhysRevLett.103.150502,morales2025quantumlinearsolverssurvey}.

\section{Derivative Estimation}\label{sec:derivative-estimation}

The protocol for estimating the matrix pair $(\mathbf{J},\mathbf{R})$ in \cref{alg:super-Krylov} requires a quantum computer to estimate the probabilities
\begin{equation}\label{eq:z-der-est}
    z(t) = \left|\bra{v} U(t) \ket{v} \right|^2
\end{equation}
at the $D$ timepoints,
\begin{equation}\label{eq:msmts}
    t\in \mathcal{T}=\{0=t_1<t_2<...<t_D=T\}
\end{equation}
with $T=m\mathrm{d}t$ for some Krlyov dimension $m$ and timestep $\mathrm{d}t$. In practice, the measurements of \eqref{eq:z-der-est} are corrupted by errors due to Trotterization, finite-shot sampling, and hardware noise in the device. Denote these noisy measurements by
\begin{equation}\label{eq:measurements}
    y_i = z(t_i) + \eta_i, \quad i=1,...,D.
\end{equation}
Given the noisy measurements $y_i$, we wish to construct a smooth estimate of each function $z(t)$ and $\frac{d}{dt}z(t)$ over $t\in\Omega \coloneqq [0,T]$. We propose a \textit{state-space} \cite{optimalControlMilanese,optimalEstimationMilanese} approach for this purpose. The approach makes the following key assumptions:
\begin{enumerate}
    \item The first $M$ derivatives of the signal $z$ are continuously differentiable over $\Omega$, for some $M\geq2$.
    \item The noise $\eta$ is bounded ($\|\eta\|_2<\infty$).
\end{enumerate}
We emphasize that we do not assume \textit{a priori} knowledge of the statistical distribution of $
\eta$, and so our estimate can be applied irrespective of the noise description of the quantum device.

By the smoothness assumption, we may define the following function $x:\Omega\subseteq\mathbb{R}\to\mathbb{R}^M$, with components comprising the true signal $z$ and its first $M$ derivatives:
\begin{equation}\label{eq:def-state-vector}
    x_k ({\gamma}) = \frac{d^k}{dt^k} z(t), \quad k=0,...,M-1.
\end{equation}
Denote the initial vector $x_{in}\coloneqq x(0)$. \cref{eq:def-state-vector} can be written equivalently as
\begin{equation}\label{eq:x-diff}
    \frac{d}{dt} x(t) = A x(t) + f(t),
\end{equation}
where $A$ is the $M\times M$ matrix
\begin{equation}\label{eq:defn A f}
    A \coloneqq
    \begin{pmatrix}
        0 & 1 & 0 & \hdots & 0 \\
        0 & 0 & 1 & \hdots & 0 \\
        \vdots & & \vdots & \ddots & \vdots \\
        0 & 0 & 0 & \hdots & 1 \\
        0 & 0 & 0 & \hdots & 0
    \end{pmatrix}
    ,
\end{equation}
and $f$ is the $M$-vector function
\begin{equation}
    f(t)\coloneqq
    \dfrac{d^{M-1}}{dt^{M-1}} x_{0} (t) \,\,
    \begin{pmatrix}
        0, & \hdots\,\, , & 0, & 1 
    \end{pmatrix}
    ^T
    .
\end{equation}
Similarly, the measurements in \eqref{eq:msmts} may be written with respect to $x$ as
\begin{equation}\label{eq:msmt-x}
    y_i = C x(t_i) + \eta_i, \quad i=1,...,D,
\end{equation}
where $C=(1,0,...,0)^T\in\mathbb{R}^M$. \cref{eq:x-diff,eq:msmt-x} constitute the \textit{state-space representation} of the true vector $x$. 

We obtain an estimate $\widehat{x}$ for the true vector $x$ as the solution of the following minimization problem:
\begin{align}\label{eq:min problem}
    \boxed{\widehat{x} \coloneqq \argmin_{{w}\in \mathcal{D}(L)} \left( \sum_{i=1}^D \left(y_i - C w(t_i)\right)^2 + \theta \int_\Omega (Lw(s))^2 ds \right)}
\end{align}
for $\theta>0$. Here $L$ is the differential operator $L  = \frac{d}{dt} - A$,  equipped the initial condition $x(0) = x_{in}$, with domain denoted $\mathcal{D}(L)$. Thus, \eqref{eq:x-diff} may be written as $Lx = f$. 

The minimization problem in \eqref{eq:min problem} admits the following interpretation. The second term of the objective function in \eqref{eq:min problem} can be expanded as
\begin{equation}\label{eq:lx-expanded}
\begin{aligned}
    \int_\Omega (Lw(s))^2 ds =& \sum_{l=0}^{M-2}\int_\Omega  \left( \frac{d}{dt} w_l (t) - w_{l+1}(t) \right)^2 dt \\ 
    &+ \int_\Omega \left(\frac{d}{dt} w_{M-1} (t) \right)^2 dt.
\end{aligned}
\end{equation}
Thus, for large $\theta$, the term in \eqref{eq:lx-expanded} dominates the objective function, which, informally, ensures that the structure of the estimate $\widehat{x}$ resembles that of the true vector $x$ in \eqref{eq:def-state-vector}, i.e., that the $k$-th component of $\widehat{x}$ is a valid approximation of the $(k-1)$-th derivative of $z$. In the limit $\theta\to\infty$, by \eqref{eq:lx-expanded}, the estimate $\widehat{x}$ satisfies
\begin{equation}\label{eq:xhat-component-convergence}
     \widehat{x}_{l+1}(t) \,{\to}\, \frac{d^l}{dt^{l}} \widehat{x}_0(t) \quad \Rightarrow \quad \frac{d}{dt} \widehat{x}_{M-1}(t) \,{\to}\, \frac{d^{M-1}}{dt^{M-1}} \widehat{x}_0(t),
\end{equation}
where convergence is defined with respect to the $\mathcal{L}_2$-norm. In turn, the $(M-1)$-th derivative of $\widehat{x}_0$ is penalized, resulting in a smoother, under-fitted estimate. On the other hand, for small $\theta$, the first term of the objective function dominates, imposing that the estimate $C\widehat{x}(t) = \widehat{x}_0(t)$ is a good fit to the datapoints $t\in\mathcal{T}$. In the limit $\theta\to0$, this would result in over-fitting of the estimate $\widehat{x}$, assuming $\|\eta\|_2\neq0$.

\subsection{Minimax formulation}\label{subsec:minimax-formulation}

\textit{Notation:} We introduce the following function spaces, each equipped with its canonical inner product and norm:
\begin{itemize}
    \item $\mathcal{L}^2(\Omega) \coloneqq \left\{h:\mathbb{R}\to\mathbb{R}| \, \int_\Omega h^2(s)\,ds < \infty \right\}$.
    \item $\mathcal{W}^{1,2}(\Omega) \coloneqq \left\{h\in\mathcal{L}^2(\Omega) | \,\frac{d}{dt}h(t)\in\mathcal{L}^2(\Omega) \right\}$.
    \item $\mathcal{V}_M^2(\Omega) \coloneqq \left\{h:\mathbb{R}\to\mathbb{R}^M |  {h}_0,..., {h}_{M-1} \in \mathcal{W}^{1,2}(\Omega) \right\}$, with shorthand $\mathcal{V}\equiv \mathcal{V}_M^2(\Omega)$.
\end{itemize}

In this section, we formulate the minimization problem in \eqref{eq:min problem} as a ``minimax'' problem. This allows us to quantify the minimax error, or ``worst-case'' error, in the corresponding estimate. We then show that the minimax estimate and error can each be expressed as the solution of an ordinary differential equation (ODE), providing a means of computing each quantity analytically.

We assume the following bound on the noise $\eta$ and the higher-order derivative of the true signal $z$:
\begin{equation}\label{eq:theta-bound0}
    \theta_2\sum_{i=1}^D \eta_i^2 + \theta_1 \left\|  \frac{d^{M-1}}{dt^{M-1}} z(t) \right\|_{\mathcal{L}^2}^2\leq 1
\end{equation}
for some $\theta_1,\theta_2>0$. By the smoothness and noise assumptions introduced in the previous section, it is always possible to choose $\theta_1,\theta_2$ such that \eqref{eq:theta-bound0} is satisfied.  Notice, by \eqref{eq:measurements} and \eqref{eq:x-diff}, that \eqref{eq:theta-bound0} can be expressed in terms of $x$ as
\begin{equation}\label{eq:theta-bound}
    \theta_2 \sum_{i=1}^D \left(y_i - C x(t_i)\right)^2 + \theta_1 \|Lx\|_\mathcal{V}^2 \leq 1.
\end{equation}
It was shown in \cite{Zhuk_2010} that the solution $\widehat{x}$ of \eqref{eq:min problem}, with regularization parameter taken as $\theta=\theta_1/\theta_2$ for $\theta_1,\theta_2$ satisfying \eqref{eq:theta-bound}, can be written equivalently as the solution of the following minimax optimization problem: 

\begin{widetext}
For all $l\in\mathcal{V}$,
\begin{equation}\label{eq:general minimax problem}
    \max_{{w} \in \mathcal{G}_y} \left| \big({l},  \widehat{{x}} - {w}\big)_{\mathcal{V}} \right| = \min_{{q} \in \mathcal{G}_y} \max_{{w} \in \mathcal{G}_y} \left| \big({l},  {q} - {w}\big)_{\mathcal{V}} \right|
\end{equation}
where $\mathcal{G}_y$ is the set of \textit{admissible solutions}
\begin{equation}\label{eq:defn bounding space}
    \mathcal{G}_y \coloneqq \left\{w\in\mathcal{D}(L): 
    \theta_2\sum_{i=1}^D \left(y_i - C w(t_i)\right)^2 + \theta_1 \| Lw \|_\mathcal{V}^2 \leq 1 \right\}.
\end{equation}
The vector $\widehat{x}$ in \eqref{eq:general minimax problem} is called a \textit{minimax estimate} of $x$, with associated \textit{minimax error},
\begin{equation}\label{eq:general minimax error}
    \Upsilon_{\widehat{x}}(l)\coloneqq \max_{{w} \in \mathcal{G}_y} \left| \big({l},  \widehat{{x}} - {w}\big)_{\mathcal{V}} \right|.
\end{equation}
\cref{eq:general minimax error} represents the ``worst-case" estimation error over all admissible solutions.

\end{widetext}

Let $\delta(\cdot)$ be the generalized Dirac delta function and let $e_k$ be the $k$-th computational vector. The following example shows how $\Upsilon_{\widehat{x}}$ may be used to upper bound the estimation error on each $\widehat{x}_k$.

\begin{example}\label{eg:minimax-error}
    Take $l(t)\coloneqq \delta(t - t^*)e_k$ for some $t^*\in\Omega$. Then, \eqref{eq:general minimax error} reads
    \begin{equation}
    \begin{split}
    \Upsilon_{\widehat{x}}(l) &= \max_{{w} \in \mathcal{G}_y} \left| \big(\phi(t - t^*)e_k,  \widehat{{x}} - {w}\big)_{\mathcal{V}} \right| \\
    &=  \max_{w \in \mathcal{G}_y} \left| \widehat{x}_k(t^*) - w_k(t^*) \right| \\
    &\geq \left| \widehat{x}_k(t^*) - x_k(t^*) \right| \label{eq:component-wise-upper-bound}
    \end{split}
    \end{equation}
    where the last line follows since $x\in\mathcal{G}$. That is, $\Upsilon_{\widehat{x}}(l)$ upper bounds the estimation error on $\widehat{x}_k$ at the time $t^*$.
\end{example}

The following theorem shows that minimax estimate \eqref{eq:general minimax problem} and corresponding minimax error \eqref{eq:general minimax error} can each be expressed as the solution of a coupled first-order ODE.

\begin{theorem}
\label{thm:full minimax expression}
    Let $\phi_s(t)=\delta(t-t_s)$. The solution $\widehat{{x}}$ of the minimax problem \eqref{eq:general minimax problem} solves
    \begin{equation}\label{eq:minimax-sol-ODE}
    \begin{split}
        \frac{d}{dt} \widehat{x}(t) &= A \widehat{x}(t) + \frac{1}{\theta_1} e_{M-1} e_{M-1}^T b(t), \\
        \frac{d}{dt} b(t) &= - A^T b(t) - \theta_2 \sum_{s=1}^{D} \phi_s(t) e_0 \left( y_s - (\phi_s e_0, \widehat{x})_\mathcal{V} \right),
    \end{split}
    \end{equation}
    for some $b\in\mathcal{V}$, with boundary conditions $\widehat{{x}}(0)=x_{in}$ and ${b}(T) = 0$. Furthermore, for $l\in\mathcal{V}$, the corresponding minimax error \eqref{eq:general minimax error} in $\widehat{x}$ is
    \begin{equation}
        \Upsilon_{\widehat{ x}}(l) = \sqrt{\left(l, \widehat{p}\right)_\mathcal{V}}
    \end{equation}
    where $\widehat{p}\in\mathcal{V}$ solves
    \begin{equation}\label{eq:minimax-err-ODE}
    \begin{split}
        \frac{d}{dt} \widehat{p}(t) &= A \widehat{p}(t) + \frac{1}{q} {e}_{M-1} e_{M-1}^T g(t), \\
        \frac{d}{dt} g(t) &= -A^T g(t) - l(t) + \theta_2 \sum_{s=1}^{D} ( \phi_s e_0, \widehat{p})_\mathcal{V} \, \phi_s(t)\,  e_0,
    \end{split}
    \end{equation}
    for some $g\in\mathcal{V}$, with boundary conditions $\widehat{p}(0) = 0$ and $g(T) = 0$.
\end{theorem}

We emphasize that, given the regularization parameter in \eqref{eq:min problem} is taken as $\theta=\theta_1/\theta_2$ such that \eqref{eq:theta-bound} holds, then the solution of \eqref{eq:min problem} and the solution of \eqref{eq:general minimax problem} are completely equivalent. Thus, the interpretation of the solution $\widehat{x}$ from the previous subsection holds, while we may use the coupled ODEs in \eqref{eq:minimax-sol-ODE} and \eqref{eq:minimax-err-ODE} to analytically compute the estimate $\widehat{x}$ and its worst-case error.

In practice, the parameters $\theta_1$ and $\theta_2$ in \eqref{eq:theta-bound0} are chosen such that 
\begin{equation}\label{eq:choose-theta}
    \theta_2 \|\eta\|_2^2 \leq \frac{1}{2} \quad \text{and} \quad \theta_1 \left\|  \frac{d^{M-1}}{dt^{M-1}} z(t) \right\|_{\mathcal{L}^2}^2 \leq \frac{1}{2}.
\end{equation}
The noise $\eta$ is estimated from the specifications of the quantum computer. The integral of $\frac{d^{M-1}}{dt^{M-1}} z(t)$ is estimated by analytically differentiating the expression for $z$ in \eqref{eq:z-der-est} and then using numerical methods to estimate the integral in \eqref{eq:choose-theta} (e.g., using Gaussian quadrature \cite{suli2003introduction}).

As the magnitude $\left|\frac{d}{dt}z(t)\right|$ increases, a higher sampling resolution will be required to accurately capture the behavior of the functions $z$ and $\frac{d}{dt}z(t)$. In \cref{subsec-app:derivative-upper-bound}, we show that, if the Hamiltonian takes the form $H=\sum_{i=1}^K\alpha_i H_i$ for some positive integer $K$, coefficients $|\alpha_i|\leq1$ and local Pauli operators $H_i$, then
\begin{equation}\label{eq:dzdt-upper-bound}
    \left|\frac{d}{dt}z(t)\right| \leq 2 m K.
\end{equation}
For example, if the Hamiltonian has nearest neighbor interactions, $H=\sum_{i=1}^{n-1}\alpha_i H_i$, then $K=n-1$, and $\left|\frac{d}{dt}z(t)\right|$ scales linearly with $n$ in the worst-case.

The following lemma (proof in \cref{sec-app:derivative-estimate-convergence}) shows that the estimate $\widehat{x}$ converges to the exact solution $x$ in the limits of zero noise and infinitely many datapoints.

\begin{lemma}\label{lemma:derivative-convergence}
    Take $\widehat{x}$ as defined in \eqref{eq:min problem}. In the limits $\|\eta\|_2\to0$ and $D\to\infty$, we have
    \begin{equation}\label{eq:x-convergence-lemma}
        \left\| \widehat{x} - x \right\|_\mathcal{V} \to 0.
    \end{equation}
\end{lemma}

\begin{figure}
    \centering
    \includegraphics[width=0.95\linewidth]{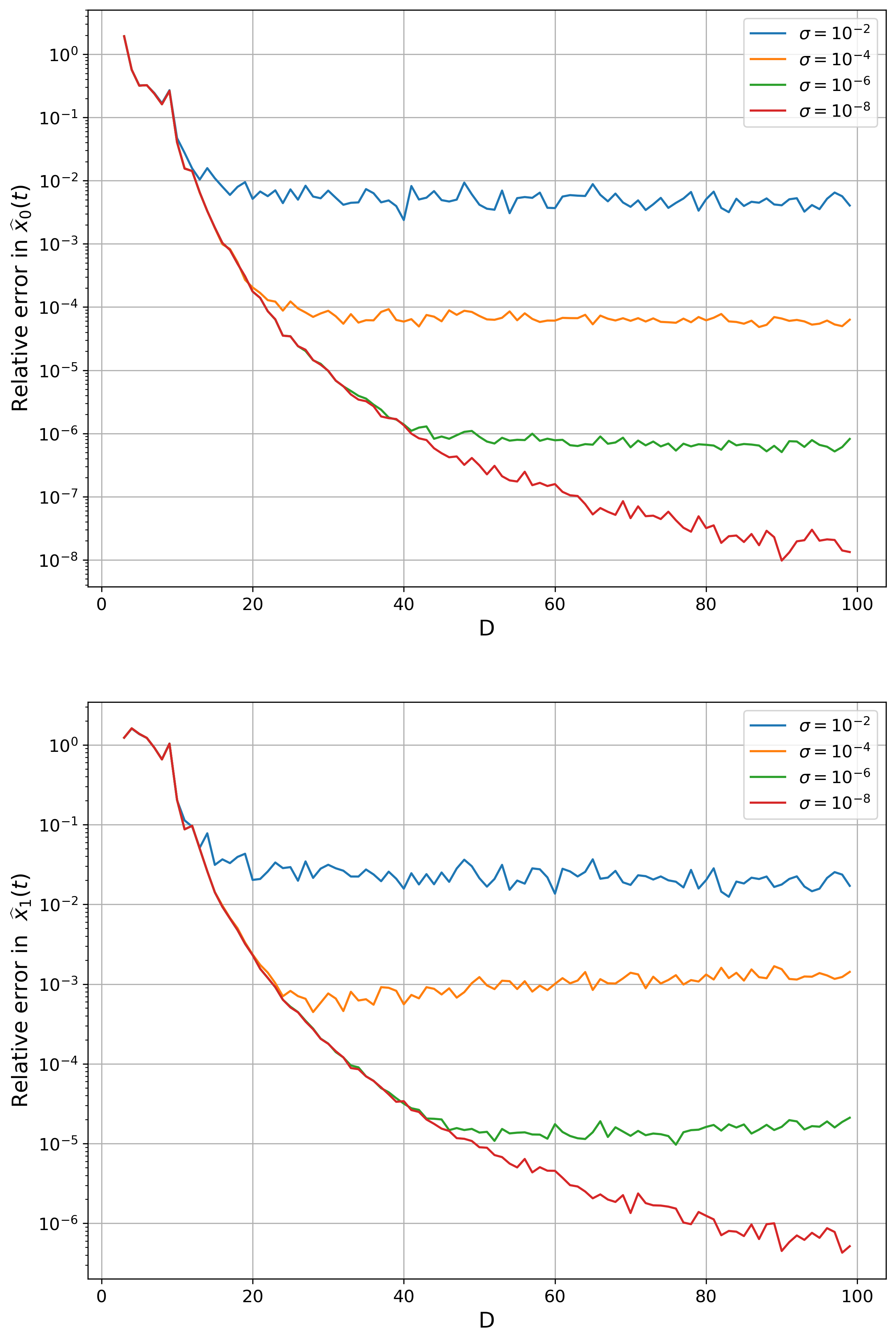}
    \caption{Convergence of the state-space estimate $\widehat{x}$ for the test function $z(t) = \cos(t) \sin(3t)$. We take $M=3$, so ${f(t)=\frac{d^2}{dt^2}z(t)}$ in \eqref{eq:x-diff}, and consider Gaussian noise $\eta$ with width $\sigma$. The top panel shows the relative error in $\widehat{x}_0(t)$ with respect to $z(t)$, averaged over $t\in \Omega = [0,2\pi]$. The bottom panel shows the relative error in $\widehat{x}_1(t)$ with respect to $\frac{d}{dt}z(t)$, again averaged over $t\in \Omega = [0,2\pi]$.}
    \label{fig:derivative-estimate}
\end{figure}

\cref{lemma:derivative-convergence} implies that each estimate $\widehat{x}_k$ converges to the true derivative $\frac{d^k}{dt^k} z(t)$ in the limits $\|\eta\|_2\to0$ and $D\to\infty$. A numerical demonstration of the convergence rate is presented in \cref{fig:derivative-estimate}.

\vspace{5mm}

\section{Krylov error analysis}\label{sec:error-analysis}

In this section, we derive an upper bound on the projection error on the super-Krylov estimate of $\Delta_0= \lambda_0 - \lambda_{N-1}$. We begin by proving that the super-Krylov estimate of $\Delta_0$ can be equivalently obtained via a standard KQD projection method. This allows us to call upon the existing error analysis in \cite{epperly2021subspacediagonalization,kirby2024analysis}, and thus derive an upper bound on the projection error.

Let $L:\mathcal{H}^{\otimes 2} \to \mathcal{H}^{\otimes 2}$ be the operator
\begin{align}\label{eq:def-L}
    L \coloneqq \overline{H}\otimes I - I \otimes H 
\end{align}
where $\overline{H}$ denotes the complex conjugation of $H$. Here $L$ has eigenvalue equation
\begin{equation}
    L\overline{\ket{u_q}} \otimes \ket{u_p} = (\lambda_p - \lambda_q) \overline{\ket{u_q }} \otimes \ket{u_p}.
\end{equation}
In particular, $L$ has the same spectrum as the super-operator $\mathfrak{L}$ in \eqref{eq:def-J}, with minimum eigenvalue $\lambda_0-\lambda_{N-1}$ and corresponding eigenvector $\overline{\ket{u_N}}\otimes \ket{u_1}$. If we define the \textit{column-major vectorization} operator \cite{hornjohnson1991matrixanalysis}, for $X \in \mathcal{M}_N$, by
\begin{equation}
    \textup{vec}(X)  \coloneqq  \sum_{j=0}^{N-1} |j \rangle \otimes X |j \rangle \in \mathbb{C}^{N^2} \,\,,
\end{equation}
then the eigenvectors of $L$ can be related to the eigenoperator of $\mathfrak{L}$ via
\begin{equation}
    \overline{|u_q \rangle} \otimes |u_p \rangle = \text{vec} (|u_p\rangle \langle u_q|).
\end{equation}

The following Lemma (proof in \cref{subsec-app:proof-of-lemma-L}) connects the super-Krylov method to a Galerkin projection scheme for the operator $L$.

\begin{lemma}\label{lemma:L-projection}
    Reinstate the notation of \cref{sec:super-krylov-method}. Let $\mathcal{Q}$ be the $m$-dimensional subspace of $\mathcal{H}^{\otimes 2}$ defined by
    \begin{equation}\label{eq:def-krylov-space-Q}
        \mathcal{Q} \coloneq \text{span} \left\{ e^{-\imath j L t} \overline{\ket{v}} \otimes \ket{v}  \right\}, \quad j=0,...,m-1.
    \end{equation}
    Consider the Galerkin projection scheme:
    \begin{equation}\label{eq:L-galerkin-proj}
        \boxed{\text{Find ${\eta} \in \mathbb{C}$ and ${w} \in \mathcal{Q}$ s.t. $L{w} - {\eta} {w}\perp\mathcal{Q}$.}}
    \end{equation}
    The quantities $\eta$ in \eqref{eq:L-galerkin-proj} solve the generalized eigenvalue problem
    \begin{equation}\label{eq:equiv-proj-problem}
        \mathbf{J}(t) a = \eta \mathbf{R}(t) a
    \end{equation}
    where $\mathbf{J}(t)$ and $\mathbf{R}(t)$ are as given in \eqref{eq:Jt-Rt-proj-prob}.
\end{lemma}

By \cref{lemma:L-projection}, the super-Krylov energy estimate in \eqref{eq:J-galerkin-proj} can be obtained equivalently via the KQD method with projection space $\mathcal{Q}$ spanned by real-time evolutions of the operator $L$. Thus, we may call on the existing KQD analysis \cite{epperly2021subspacediagonalization} to produce an error bound on our estimate of $\Delta_0$. We restate Theorem 3.1. from \cite{epperly2021subspacediagonalization} in the present notation, which provides an upper bound on the energy estimation error in the KQD method.

\begin{theorem}[Theorem 3.1. in \cite{epperly2021subspacediagonalization}, expressed asymptotically]\label{thm:asymptotic error bound}
    Reinstate the notation of \cref{sec:super-krylov-method}. Set 
    \begin{equation}\label{eq:timestep-cond}
        \mathrm{d}t = \pi / 2(\lambda_0 - \lambda_{N-1}).
    \end{equation}
    Let $\big(\widehat{\mathbf{J}
    },\widehat{\mathbf{R}}\big)$ be a Hermitian approximation of the pair $(\mathbf{J},\mathbf{R}) \equiv (\mathbf{J}(\mathrm{d}t),\mathbf{R}(\mathrm{d}t))$, and let 
    \begin{equation}
    \label{eq:chi_def}
        \omega \coloneqq \|\widehat{\mathbf{R}}-\mathbf{R}\,\|_2+\frac{\|\widehat{\mathbf{J}}-\mathbf{J}\,\|_2}{\| L\|_2}
    \end{equation}
    be a measure of the noise. Let $\widehat{\Delta}_0$ be the minimum eigenvalue of the $\epsilon$-thresholded pair $\big(\widehat{\mathbf{J}
    },\widehat{\mathbf{R}}\big)$ for $\epsilon = O(\omega^{2/3})$. Define the initial overlap
    \begin{equation}\label{eq:initial-overlap-expression}
        \gamma_0 \coloneqq \braket{v}{u_0}\braket{u_{N-1}}{v},
    \end{equation}
    and assume that
    \begin{equation}\label{eq:overlap-cond}
        | \gamma_0| > 2 ( \epsilon + \| \widehat{\mathbf{R}} - \mathbf{R} \| ).
    \end{equation}
    Then
    \begin{equation}
    \label{eq:krylov_error_bound_epperly}
        \frac{\left|\widehat{\Delta}_0-\Delta_0\right|}{|\Delta_0|}\le O\left(\frac{\omega^{2/3} + \left(1+\frac{\Delta_1-\Delta_0}{|\Delta_0|}\right)^{-2m}}{|\gamma_0|^2}\right).
    \end{equation}
\end{theorem}

We omit a proof of \cref{thm:asymptotic error bound}, as it is obtained by simply applying Theorem 3.1. of \cite{epperly2021subspacediagonalization} to the KQD scheme in \eqref{eq:L-galerkin-proj}. From \eqref{eq:krylov_error_bound_epperly}, one can see that the error in the estimate $\widehat{\Delta}_0$ decays exponentially quickly with the Krylov dimension $m$, to a minimum value that is proportional to the noise rate $\omega$.

The two key differences between Theorem 3.1. in \cite{epperly2021subspacediagonalization} and its reformulation in \cref{thm:asymptotic error bound} lie in the timestep value and the initial overlap. For the KQD method applied to an operator $H$, the analysis in \cite{epperly2021subspacediagonalization} employs a timestep value of $\mathrm{d}t = \pi / R(H)$ to obtain the upper error bound, where $R(\cdot)$ denotes the spectral range. The KQD method in \eqref{eq:L-galerkin-proj} uses time evolutions of the operator $L$, which yields $\mathrm{d}t = \pi / R(L) = \pi / 2 R(H)$, so the super-Krylov method benefits from a halved time evolution. Condition \eqref{eq:overlap-cond} implies that the initial state $\ket{v}$ must have a non-zero overlap between the ground state $\ket{u_0}$ and the most-excited state $\ket{u_{N-1}}$ of $H$. The latter condition is an additional requirement to the standard KQD method, where one only requires a non-zero overlap with the ground state. In \cref{sec:experiments}, we provide an example of how such an initial state may be constructed.

The condition $\epsilon = O(\omega^{2/3})$ in \cref{thm:asymptotic error bound} can be interpreted informally as ensuring eigenvectors that are compatible with the noise level are truncated out after thresholding. In practice, it is typically found that $\epsilon = O(\omega)$ is a preferable choice \cite{kirby2024analysis}, and so we employ this practical choice in our experiments in \cref{sec:experiments}. Finally, the condition \eqref{eq:overlap-cond} can be interpreted as requiring that the initial overlap is not overwhelmed by the error from thresholding and the noise. For a detailed discussion of \cref{thm:asymptotic error bound} and the corresponding proof, we refer the reader to \cite[Section 3.1.]{epperly2021subspacediagonalization}.

\section{Numerical Experiments}\label{sec:experiments}

We benchmark the super-Krylov method on the spin-1/2 nearest-neighbor transverse-field Ising (TFI) Hamiltonian
\begin{equation}\label{eq:tfi hamiltonian}
    H =  - J \sum_{i=0}^{n-2}  \, \sigma^x_i \sigma^x_{i+1}
    - g \sum_{j=0}^{n-1} \, \sigma^z_j
\end{equation}
with couplings $J,g\in\mathbb{R}$. Take $W=(\sigma^x \sigma^z)^{\otimes n/2}$, then $\{H,W\}=0$ and so the TFI model belongs to the first class in \cref{subsec:superk-applications}. Thus, we may estimate the ground-state energy as $\widehat{\lambda}_0 \coloneqq \widehat{\Delta}_0/2$. To fulfill the overlap condition $\gamma_0\neq0$, we take the initial state as a superposition of the ground state and most-excited state of the first term of the Hamiltonian in \eqref{eq:tfi hamiltonian}. That is, we take the \textit{Greenberger–Horne–Zeilinger} (GHZ)-like state
\begin{equation}
    \ket{v} = \frac{1}{\sqrt{2}}\left( \ket{+-}^{\otimes n/2} + \ket{+}^{\otimes n} \right).
\end{equation}
For coupling values $(J,g)=(1,0.35)$, this gives an initial overlap of ${|\gamma_0|=|\braket{v}{u_0}\braket{v}{u_{N-1}}|=0.215}$.

\begin{figure}
    \includegraphics[width=0.48\textwidth]{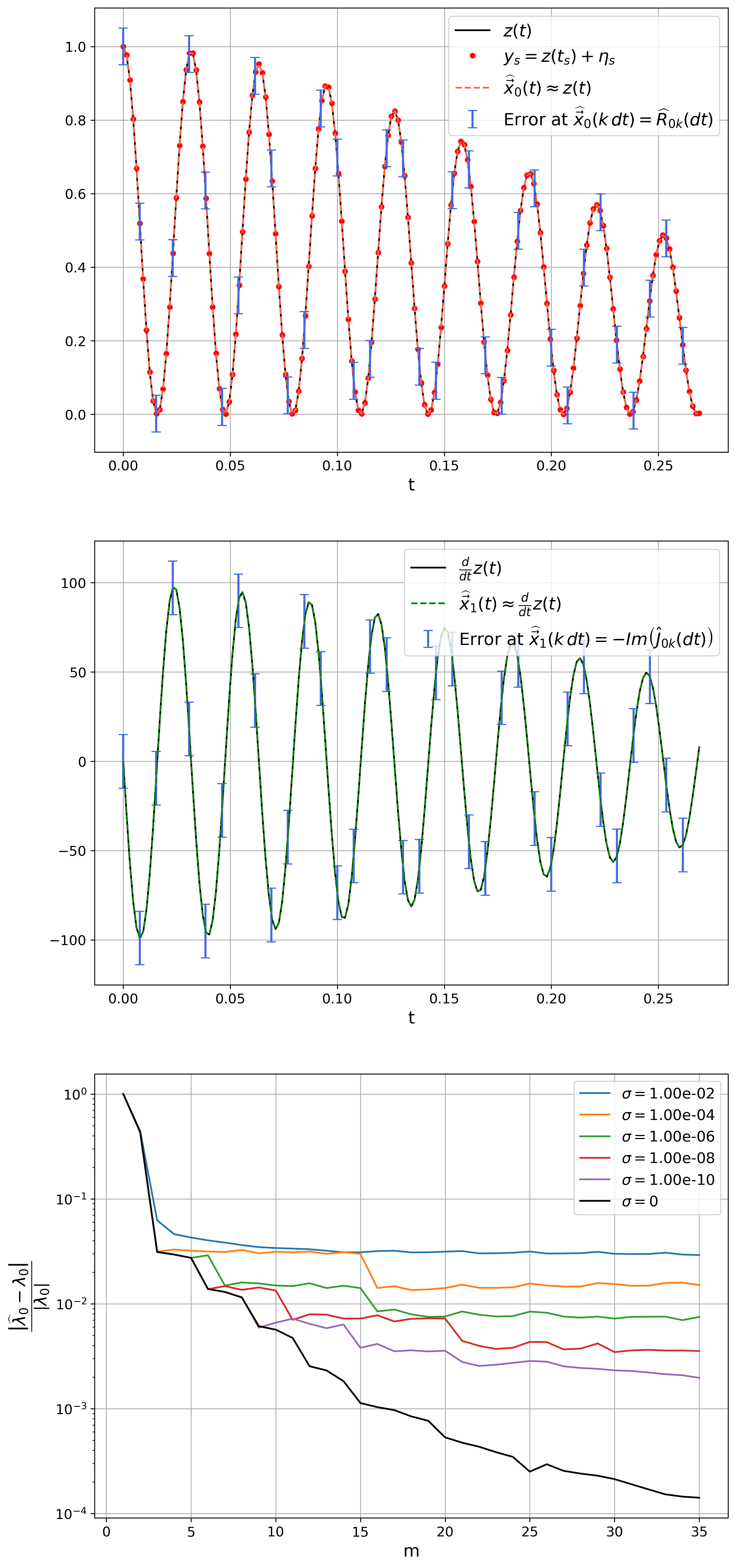}

    \caption{Numerical demonstration of the super-Krylov for the TFI model on 100 qubits. The top panel shows the exact function $z(t)$, $D=5m$ noisy measurements $y$ for $\sigma=0.1$, and function estimate $\widehat{{x}}_0(t)$. The middle panel shows the exact function $\frac{d}{dt}z(t)$ and corresponding function estimate $\widehat{ {x}}_1(t)$. The worst-case (minimax) estimation error is plotted in each case. The lower panel shows the error in the ground energy estimate $\widehat{\lambda}_0$ versus the Krylov dimension $m=35$, for various noise rates $\sigma$.}
    \label{fig:TFI}
\end{figure}

The numerical results are presented in \cref{fig:TFI}. The signal $z(t) = |\langle w | U(t) | w \rangle |^2$ is obtained using the time-evolving-block-decimation (TEBD) algorithm in TeNPy, a Python package for tensor network simulations \cite{tenpy}, with timestep $\mathrm{d}t$ taken as per \cref{thm:asymptotic error bound}. We use the density matrix renormalization group (DMRG) \cite{dmrg1,dmrg2,dmrg3,Schollwock2005} as a ground truth for the eigenpairs $(\lambda_0,\ket{u_0})$ and $(\lambda_{N-1},\ket{u_{N-1}})$.

For the state-space estimate $\widehat{x}$, we take $M=3$ and use Gaussian noise $\eta$ with widths $\sigma$. The upper panel in \cref{fig:TFI} shows the true function $z$, noisy measurements $y$ for $\sigma=0.01$, and the corresponding function estimate $\widehat{x}_0(t)$ and minimax error. The middle panel in \cref{fig:TFI} shows the true function $\frac{d}{dt} z(t)$ and the associated estimate $\widehat{{x}}_1(t)$ and minimax error. The minimax error bars represent the worst-case estimation error at the Krylov timesteps $k\mathrm{d}t$ (see \cref{eg:minimax-error}). The lower panel in \cref{fig:TFI} shows the relative error in the ground-state energy estimate $\widehat{\lambda}_0=\widehat{\Delta_0}/2$ for various levels of noise.

\section{Conclusions}

In conclusion, we have presented a novel addition to the class of ``near-term friendly" KQD methods. The projection error in our energy estimate converges exponentially quickly with the Krylov dimension, as is observed in the standard KQD approach. We also proposed a novel state-space derivative estimation algorithm for use in post-processing, and presented a proof of convergence of the method. Our numerical experiments show the capability of the method for estimating the ground-state energy of Hamiltonians exhibiting particular symmetries.

One clear direction for follow-up work would be to demonstrate the super-Krylov method on a quantum device in order to display the method's compatibility with existing quantum processors. On the theoretical side, it would be fruitful to derive an expression for the error in the state-space estimate with respect to the sampling errors and the number of datapoints. This would allow for an
end-to-end error analysis of the energy estimate. One could
also consider adaptive sampling strategies in order to reduce the number of measurements required from the quantum device.

\section*{Acknowledgements}
A.B. thanks Sergey Bravyi (IBM), Antonio Mezzacapo (IBM) and Nicola Mariella (IBM) for the helpful discussions.

\clearpage

\appendix
\onecolumngrid

\section{The super-Krylov method}\label{sec-app:super-k}

\subsection{Derivation of the projected eigenvalue problem \eqref{eq:Jt-Rt-proj-prob}}

Recall from \cref{sec:super-krylov-method} that the super-Krylov method obtains approximate eigenpairs $(\widetilde{\Delta},\widetilde{X})$ of $\mathfrak{L}$ via the Galerkin projection:
\begin{equation}\label{eq:J-galerkin-proj-apx}
    \boxed{\text{Find $\widetilde{\Delta}\in\mathbb{C}$ and $\widetilde{X} \in \mathcal{B}$ s.t. $\mathfrak{L}\widetilde{X} - \widetilde{\Delta}\widetilde{X} \perp \mathcal{B}$}}
\end{equation}
where $\mathcal{B}$ is an $m$-dimensional space in $\mathcal{M}_N$, with basis vectors $B=\{\rho_0,...,\rho_{m-1}\}$. Note that the orthogonality condition in \eqref{eq:J-galerkin-proj-apx} is defined with respect to the trace inner product. \cref{eq:J-galerkin-proj-apx} implies that
\begin{equation}\label{eq:super-k derive 1}
    \text{Tr}\bigg( \big( \mathfrak{L} \widetilde{X} - \widetilde{\Delta} \widetilde{X}  \big)^\dag \rho_k \bigg) = 0, \qquad k=0,...,m-1.
\end{equation}
Since $\widetilde{X}\in\mathcal{B}$ then $\widetilde{X}=\sum_{j=0}^{m-1} a_j \rho_j$ for some $a_j\in\mathbb{R}$. Inserting this expression for $\widetilde{X}$ into \eqref{eq:super-k derive 1} yields
\begin{equation}\label{eq:super-k derive 2}
    \sum_{j=0}^{m-1}\text{Tr}\bigg( \big(\mathfrak{L} \rho_j \big)^\dagger \rho_k   \bigg) a_j = \widetilde{\Delta} \sum_{j=0}^{m-1} \text{Tr}\big( \rho_j^\dagger \rho_k  \big) \, a_j,
\end{equation}
or equivalently, in matrix notation,
\begin{align}\label{eq:J projected eigenvalue problem apndx}
    \mathbf{J}a  = \widetilde{\Delta} \, \mathbf{R}a
\end{align}
where $\mathbf{J}_{jk} = \text{Tr} ( (\mathfrak{L}\rho_j)^\dag \, \rho_k )$ and $\mathbf{R}_{jk} = \text{Tr}(\rho_j^\dag \rho_k )$. Now, take the basis vectors
\begin{equation}\label{eq:defn-rhoj-app}
    \rho_j(t) = U^j(t) \rho \, U^{-j}(t), \quad j=0,...,m-1,
\end{equation}
for some initial state $\rho=\ketbra{v}{v}$ and time $t\in\mathbb{R}$. Since $\rho_j(t)$ are Hermitian, \eqref{eq:J projected eigenvalue problem apndx} becomes
\begin{align}\label{eq:J projected eigenvalue problem apndx-1}
    \mathbf{J}(t)a  = \widetilde{\Delta} \, \mathbf{R}(t)a
\end{align}
where $\mathbf{J}_{jk}(t) = \text{Tr} \left(\big( \mathfrak{L}\rho_j(t)\big) \, \rho_k(t) \right)$ and $\mathbf{R}_{jk}(t) = \text{Tr}(\rho_j(t) \rho_k(t) )$. 

We proceed to derive the expression for $\mathbf{R}(t)$ in \eqref{eq:R-probabilities}. Using the cyclic property of the trace, $\mathbf{R}_{jk}(t)$ can be expanded as
\begin{equation}
    \begin{split}
        \mathbf{R}_{jk}(t) &= \text{Tr}\left(\rho_j(t)\rho_k(t)\right) \\
        &= \text{Tr}\biggl(U^j(t)\ketbra{v}{v}U^{-j} U^k(t)\ketbra{v}{v}U^{-k}\biggr) \\
        &= \left| \bra{v}U^{-j} U^k(t) \ket{v}\right|^2.
    \end{split}
\end{equation}
We now derive the relation between $\mathbf{R}(t)$ and $\mathbf{J}(t)$ in \eqref{eq:derivative-relation}. Direct differentiation of $\mathbf{R}_{jk}(t)$ yields
\begin{equation}\label{eq:ddt R proof 1}
        \frac{d}{dt} \mathbf{R}_{jk}(t) = \imath \,\text{Tr}\left( \frac{d}{dt} \rho_j(t) \rho_k(t)\right) + \imath \,\text{Tr}\left(\rho_j(t) \frac{d}{dt} \rho_k(t)\right).
\end{equation}
By the Liouville-von Neumann equation \eqref{eq:liouville-von-neumann}, we have
\begin{equation}\label{eq:ddt-rhoj}
    \imath \frac{d}{dt}\rho_j(t) = j \mathfrak{L}\rho_j(t).
\end{equation}
Inserting the expression \eqref{eq:ddt-rhoj} for $\frac{d}{dt}\rho_j(t)$ into \eqref{eq:ddt R proof 1} yields
\begin{equation}\label{eq:ddt R proof 2}
    \begin{split}
        \frac{d}{dt} \mathbf{R}_{jk}(t) &= -\imath j \, \text{Tr}\left( (\mathfrak{L}\rho_j(t)) \, \rho_k(t) \right) - \imath k \, \text{Tr}\left( \rho_j(t) \, ( \mathfrak{L}\rho_k(t)) \right).
    \end{split}
\end{equation}
Notice that the second term in \eqref{eq:ddt R proof 2} can be rearranged as
\begin{equation}\label{eq:ddt R proof 3}
    \text{Tr}\left( \rho_j(t) \, ( \mathfrak{L}\rho_k(t)) \right) = - \text{Tr}\big( (\mathfrak{L}\rho_j(t)) \rho_k(t) \big).
\end{equation}
Thus \eqref{eq:ddt R proof 2} becomes 
\begin{align}\label{eq:relation between R and J}
    \frac{d}{dt} \mathbf{R}_{jk}(t) &= \imath (k-j)\, \text{Tr}\big( (\mathfrak{L}\rho_j(t)) \rho_k(t) \big) \\
    &= \imath (k-j)\, \mathbf{J}_{jk}(t).
\end{align}

\subsection{Proof of \cref{lemma:L-projection}}\label{subsec-app:proof-of-lemma-L}

We begin by stating some useful properties of the Kronecker product.

\begin{lemma}[Kronecker product properties]
\label{lemma:kronecker properties}  
    Take $A,B,C,D\in\mathbb{C}^{N\times N}$ and define the column-major vectorization operator $\text{vec}:\mathbb{C}^{N \times N} \to \mathbb{C}^{N^2}$,
    \begin{equation}
        \textup{vec}(A)  \coloneqq  \sum_{j=0}^{N-1} |j \rangle \otimes A |j \rangle.
    \end{equation}
    Define the Kronecker sum
    \begin{equation}
        A \oplus B  \coloneqq  A \otimes I + I \otimes B.
    \end{equation}
    Then:
    \begin{enumerate}
        \item $( A \otimes B ) ( C \otimes D) = AC \otimes BD$.
        \item \textup{spec}$(A\otimes B) = \left\{ \lambda_A \lambda_B : \, \lambda_A \in \textup{spec}(A), \, \lambda_B \in \textup{spec}(B) \right\}$.
        \item 
        $\textup{vec} (A C B) = (B^T \otimes A) \, \textup{vec}(C)$.
        \item $\textup{Tr}(A^\dag B) = \textup{vec}(A)^\dag \, \textup{vec}(B)$.
        \item For $\ket{x},\ket{y}\in\mathbb{C}^N$, \textup{vec}$(|x \rangle \langle y |) = \overline{|y \rangle} \otimes |x \rangle$.
        \item $e^A \otimes e^B = e^{A \oplus B}$.
    \end{enumerate}
\end{lemma}

We proceed with the proof of \cref{lemma:L-projection}.

\noindent\textbf{Lemma \ref{lemma:L-projection}.} \emph{Reinstate the notation of \cref{sec:super-krylov-method}. Let $\mathcal{Q}$ be the $m$-dimensional subspace of $\mathcal{H}^{\otimes 2}$ defined by
\begin{equation}\label{eq:L-subspace}
    \mathcal{Q} \coloneq \text{span} \left\{ e^{-\imath j L t} \overline{\ket{v}} \otimes \ket{v}  \right\}, \quad j=0,...,m-1.
\end{equation}
Consider the Galerkin projection scheme:
\begin{equation}\label{eq:L-galerkin-proj-app}
    \boxed{\text{Find ${\eta} \in \mathbb{C}$ and ${w} \in \mathcal{Q}$ s.t. $L{w} - {\eta} {w}\perp\mathcal{Q}$.}}
\end{equation}
The quantities $\eta$ in \eqref{eq:L-galerkin-proj-app} solve the generalized eigenvalue problem
\begin{equation}\label{eq:equiv-proj-problem-app}
    \mathbf{J}(t) a = \eta \mathbf{R}(t) a
\end{equation}
where $\mathbf{J}(t)$ and $\mathbf{R}(t)$ are as given in \eqref{eq:J projected eigenvalue problem apndx-1}.
}
\begin{proof}
    Following the standard KQD method \cite{epperly2021subspacediagonalization}, the Galerkin projection scheme \eqref{eq:L-galerkin-proj-app} yields the generalized eigenvalue problem
    \begin{equation}\label{eq:equiv-proj-problem-2}
        \mathbf{L}(t) c = \eta \mathbf{T}(t) c
    \end{equation}
    where the $m\times m$ matrices $\mathbf{L}(t)$ and $\mathbf{T}(t)$ have entries
    \begin{align}
        \mathbf{L}_{jk}(t) &= \overline{\bra{v}} \otimes \bra{v} \left( e^{ \imath  j L t} L e^{-\imath  k L t} \right) \overline{\ket{v}} \otimes \ket{v}, \label{eq:proj-L-entries}  \\
        \mathbf{T}_{jk}(t) &= \overline{\bra{v}} \otimes \bra{v} \left( e^{ \imath  j L t} e^{-\imath  k L t} \right) \overline{\ket{v}} \otimes \ket{v}. \label{eq:proj-T-entries}
    \end{align}
    We will show that $(\mathbf{L}(t),\mathbf{T}(t)) = (\mathbf{J}(t),\mathbf{R}(t))$, for $\mathbf{J}(t)$ and $\mathbf{R}(t)$ as given in \eqref{eq:J projected eigenvalue problem apndx-1}. By definition, $L=\overline{H} \oplus (-H)$. Then, using \cref{lemma:kronecker properties}, property 6, the time evolution of $L$ can be written
    \begin{align}\label{eq:lemma-L-proof-1}
        e^{-\imath j L t} &= e^{ij \left(\overline{H} \oplus (-H)\right) t}  \\
        &= e^{ ij \overline{H} t} \otimes e^{ -ij H t}.
    \end{align}
    Using the fact that $H$ is Hermitian, so $\overline{H}=H^T$, and $U(t)=e^{- \imath H t}$, we get $e^{\imath j \overline{H}t} = \left(U^{-j}(t)\right)^T$. Then \eqref{eq:lemma-L-proof-1} becomes
    \begin{align}\label{eq:lemma-L-proof-2}
        e^{-\imath j L t} &= \left(U^{-j}(t)\right)^T \otimes U^{j}(t).
    \end{align}
    By \cref{lemma:kronecker properties}, property 5, the initial state in \eqref{eq:L-subspace} can also be written as
    \begin{equation}\label{eq:lemma-L-proof-3}
        \overline{\ket{v}} \otimes \ket{v} = \textup{vec} (\ketbra{v}{v}).
    \end{equation}
    Using \eqref{eq:lemma-L-proof-2} and \eqref{eq:lemma-L-proof-3}, we have
    \begin{align}\label{eq:lemma-L-proof-4}
        e^{-\imath j L t} \overline{\ket{v}} \otimes \ket{v} &= \left( \left(U^{-j}(t)\right)^T \otimes U^{j}(t) \right) \textup{vec} (\ketbra{v}{v}) \\
        &= \textup{vec} \left( U^{j}(t) \ketbra{v}{v} U^{-j}(t) \right) \\
        &= \textup{vec} \left( \rho_j(t) \right),
    \end{align}
    where the last two lines follow from \cref{lemma:kronecker properties}, property 3, and the definition of the Hermitian matrices $\rho_j(t)$ in \eqref{eq:defn-rhoj-app}, respectively. Plugging the expression \eqref{eq:lemma-L-proof-4} back into the expression \eqref{eq:proj-T-entries} for $\mathbf{T}$ yields
    \begin{align}\label{eq:lemma-L-proof-5}
        \mathbf{T}_{jk} (t) &= \overline{\bra{v}} \otimes \bra{v} \left( e^{ \imath  j L t} e^{-\imath  k L t} \right) \overline{\ket{v}} \otimes \ket{v} \\
        &= \textup{vec} \left( \rho_j(t) \right) ^\dagger \, \textup{vec} \left( \rho_k(t) \right)
    \end{align}
    By \cref{lemma:kronecker properties}, property 4, the entries in \eqref{eq:lemma-L-proof-5} become
    \begin{align}\label{eq:lemma-L-proof-6}
        \mathbf{T}_{jk} (t) &= \textup{Tr} \left(\rho_j(t) \rho_k(t)\right) \\
        &= \mathbf{R}_{jk}(t),
    \end{align}
    as desired.

    Now, for the matrix $\mathbf{L}(t)$, by \eqref{eq:lemma-L-proof-4} and the definition of $L$, we can expand the entries as
    \begin{align}
        \mathbf{L}_{jk}(t) &= \overline{\bra{v}} \otimes \bra{v} \left( e^{ \imath  j L t} L e^{-\imath  k L t} \right) \overline{\ket{v}} \otimes \ket{v} \\
        &= \textup{vec} \left( \rho_j(t) \right)^\dagger \left(\overline{H}\otimes I - I \otimes H \right) \textup{vec} \left( \rho_k(t) \right) \\
        &=  \textup{vec} \left( \rho_j(t) \right)^\dagger \left( \overline{H}\otimes I \right) \textup{vec} \left( \rho_k(t) \right) - \textup{vec} \left( \rho_j(t) \right)^\dagger \left(I \otimes H \right) \textup{vec} \left( \rho_k(t) \right). \label{eq:lemma-L-proof-7}
    \end{align}
    By \cref{lemma:kronecker properties}, property 3,
    \begin{align}\label{eq:lemma-L-proof-8}
        \mathbf{L}_{jk}(t) =  \textup{vec} \left( \rho_j(t) \right)^\dagger \textup{vec} \left( I \rho_k(t) \overline{H}^T \right) - \textup{vec} \left( \rho_j(t) \right)^\dagger \textup{vec} \left( H \rho_k(t) I \right).
    \end{align}
    Since $H$ is Hermitian, $\overline{H} = H^T$, so $\overline{H}^T = H$. Then, using \cref{lemma:kronecker properties}, property 4, \eqref{eq:lemma-L-proof-8} becomes
    \begin{align}
        \mathbf{L}_{jk}(t) &= \textup{Tr}\left( \rho_j(t) \rho_k(t) H \right) - \textup{Tr}\left( \rho_j(t) H \rho_k(t) \right) \\
        &= \textup{Tr}\left( [H,\rho_j(t)] \rho_k(t) \right) \\
        &= \mathbf{J}_{jk}(t).
    \end{align}
    Hence, the proof is complete.
\end{proof}

\section{State-space derivative estimate}\label{sec-app:derivative-estimate}

This section focuses on the proofs of the results presented in \cref{sec:derivative-estimation}. In \cref{subsec-app:derivative-upper-bound}, we derive the upper bound on $\left|\frac{d}{dt}z\right|$ in \eqref{eq:dzdt-upper-bound}. In \cref{sec-app:derivative-estimate-convergence}, we present a proof of \cref{lemma:derivative-convergence}, and in \cref{subsec-app:minimax-thm-proof} we present a proof of \cref{thm:full minimax expression}. 

Recall the following function spaces, each equipped with its canonical inner product and induced norm:
\begin{itemize}
    \item $\mathcal{L}^2(\Omega) \coloneqq \left\{h:\mathbb{R}\to\mathbb{R}| \, \int_\Omega h^2(s)\,ds < \infty \right\}$.
    \item $\mathcal{W}^{1,2}(\Omega) \coloneqq \left\{h\in\mathcal{L}^2(\Omega) | \,\frac{d}{dt}h(t)\in\mathcal{L}^2(\Omega) \right\}$.
    \item $\mathcal{V}_M^2(\Omega) \coloneqq \left\{h:\mathbb{R}\to\mathbb{R}^M |  {h}_0,..., {h}_{M-1} \in \mathcal{W}^{1,2}(\Omega) \right\}$, with shorthand $\mathcal{V}\equiv \mathcal{V}_M^2(\Omega)$.
\end{itemize}

We recall the derivative estimation framework from \cref{sec:derivative-estimation}. The function $z:[0,\infty)\to[0,1]$ is defined by
\begin{equation}\label{eq:def-z-app}
    z(t) \coloneqq  |\langle v | U(t) | v \rangle |^2
\end{equation}
where $U(t) = e^{-\imath H t}$. Define the grid of $D$ equally spaced timepoints
\begin{equation}\label{eq:msmts-app}
    \mathcal{T}\coloneqq \{0=t_1<t_2<...<t_D=T\}
\end{equation}
where $T=m \,\mathrm{d}t$, for some Krylov dimension $m$ and timestep $\mathrm{d}t$. In step 1 of \cref{alg:super-Krylov}, the function $z$ is measured at the timepoints $t\in\mathcal{T}$, yielding the noisy measurements
\begin{equation}\label{eq:measurements-app}
    y_i  \coloneqq  z(t_i) + \eta_i, \quad i=1,...,D.
\end{equation}
In step 2 of \cref{alg:super-Krylov}, classical estimates of the functions $z$ and $\frac{d}{dt}z$ are constructed, using the noisy measurements in \eqref{eq:measurements-app}. We propose a state-space \cite{optimalControlMilanese,optimalEstimationMilanese} approach for this purpose. Our approach assumes that:
\begin{enumerate}
    \item The first $M$ derivatives of the signal $z$ are continuously differentiable over $\Omega=[0,T]$, for some $M\geq2$.
    \item The noise $\eta$ is bounded, i.e., $\|\eta\|_2<\infty$.
\end{enumerate}

By the smoothness assumption, we may define the function $x:\Omega\subseteq\mathbb{R}\to\mathbb{R}^M$, with components comprising the true signal $z$ and its first $M$ derivatives:
\begin{equation}\label{eq:def-state-vector-app}
    x_k ({\gamma}) = \frac{d^k}{dt^k} z(t), \quad k=0,...,M-1.
\end{equation}
We refer to $x$ as the \textit{state vector}. Denote the initial condition $x_{in}\coloneqq x(0)$. Here, $x$ satisfies the differential equation
\begin{equation}\label{eq:x-diff-app}
    \frac{d}{dt} x(t) = A x(t) + f(t),
\end{equation}
where $A$ is the $M\times M$ matrix
\begin{equation}
    A \coloneqq
    \begin{pmatrix}
        0 & 1 & 0 & \hdots & 0 \\
        0 & 0 & 1 & \hdots & 0 \\
        \vdots & & \vdots & \ddots & \vdots \\
        0 & 0 & 0 & \hdots & 1 \\
        0 & 0 & 0 & \hdots & 0
    \end{pmatrix}
    ,
\end{equation}
and $f$ is the $M$-vector function
\begin{equation}
    f(t)\coloneqq
    \dfrac{d^{M-1}}{dt^{M-1}} x_{0} (t) \,\,
    \begin{pmatrix}
        0, & \hdots\,\, , & 0, & 1 
    \end{pmatrix}
    ^T
    .
\end{equation}
\cref{eq:x-diff-app} can be simplified further by defining the differential operator $L:\mathcal{D}(L)\subset \mathcal{V}\to\mathcal{V}$ by
\begin{equation}\label{eq:defn L}
    L{w} \coloneqq  \frac{d}{dt}{w}(t) - A{w}(t), \qquad {w}(0)=x_{in}.
\end{equation}
Here $L$ is referred to as the \textit{state operator}. We may write \eqref{eq:x-diff-app} equivalently as
\begin{equation}\label{eq:lxf}
    L x = f.
\end{equation}

We can reformulate the measurement equation in \eqref{eq:msmts-app} in a similar way to \eqref{eq:lxf}. First, define the generalized Dirac delta functions $\phi_i(t) = \delta(t-t_i)$, centered at the timepoints $t_i$. Denote by ${e}_j$ the vector with $j$-th entry equal to $1$ and all other entries equal to $0$. We define the \textit{output operator} $\widetilde{C}:\mathcal{V}\to\mathbb{R}^D$ by
\begin{equation}\label{eq:defn-Ctilde}
    \widetilde{C}{w}  \coloneqq 
    \begin{pmatrix}
         (\phi_1 {e}_0, {w})_{\mathcal{V}} \\
         \vdots \\
        (\phi_{D} {e}_0, {w})_{\mathcal{V}}
    \end{pmatrix}
    =
    \begin{pmatrix}
        {w}_0(t_1) \\
        \vdots \\
        {w}_0(t_{D})
    \end{pmatrix}
\end{equation}
where the second equality follows since
\begin{equation}
    (\phi_j {e}_0, {w})_{\mathcal{V}} = \int_\Omega\delta(t-t_j) {e}_0^T \,\, {w}(t)dt = \int_\Omega\delta(t-t_j) {w}_0(t)dt = {w}_0(t_j)
\end{equation}
for $j=1,...,D$. The tilde on $C$ is justified by the difference in definition between the operator $\widetilde{C}$ in \eqref{eq:defn-Ctilde} and the vector $C=(1,0,...,0)^T\in\mathbb{R}^M$ in \cref{subsec:minimax-formulation}. The measurements in \eqref{eq:measurements-app} can then be written as
\begin{equation}\label{eq:Cxn}
    {y} = \widetilde{C} {x} + {\eta}.
\end{equation}
\cref{eq:lxf,eq:Cxn} form the \textit{state-space representation} of the state vector $x$.

Our estimate $\widehat{x}$ of $x$ is obtained via the following minimization problem:
\begin{align}\label{eq:min-problem-app}
    \boxed{\widehat{x} \coloneqq \argmin_{{w}\in \mathcal{D}(L)} \, \| y - \widetilde{C} w\|^2_2 + \theta \| Lw \|^2_\mathcal{V}}
\end{align}
for some regularization parameter $\theta>0$. We emphasize that the formulation in \eqref{eq:min-problem-app} is equivalent to that in \eqref{eq:min problem}. For an interpretation of the terms in \eqref{eq:min-problem-app}, see \cref{sec:derivative-estimation}.

Finally, recall from \cref{sec:derivative-estimation} that, in order to reformulate the minimization problem \eqref{eq:min-problem-app} as the minimax optimization problem \eqref{eq:general minimax problem}, we assume that the parameter $\theta$ in \eqref{eq:min-problem-app} is taken as $\theta=\theta_1/\theta_2$, where
\begin{equation}\label{eq:theta-app}
    \theta_2 \sum_{i=1}^D \left(y_i - C x(t_i)\right)^2 + \theta_1 \|Lx\|_\mathcal{V}^2 \leq 1,
\end{equation}
or equivalently, in the present notation,
\begin{equation}\label{eq:theta-app-operator}
    \theta_2 \|y - \widetilde{C}x \|_2^2 + \theta_1 \|Lx\|_\mathcal{V}^2 \leq 1.
\end{equation}

\subsection{Upper bound on $\left|\frac{d}{dt}z(t)\right|$}\label{subsec-app:derivative-upper-bound}

\begin{lemma}\label{lemma:derivative upper bound}
    Take $z$ as defined in \eqref{eq:def-z-app} and suppose the Hamiltonian $H$ takes the form $H=\sum_{i=1}^K \alpha_i H_i$, where $H_i$ are local Pauli operators and $\alpha_i\in\{w\in\mathbb{C}:\,|w|\leq 1\}$. Then, for a Krylov dimension $m$,
    \begin{equation}\label{eq:derivative upper bound}
        \left| \frac{d}{dt}z(t) \right|  \leq 2mK, \qquad \forall t\in\Omega.
    \end{equation}
\end{lemma}
\begin{proof}
    From the definition of $z$, 
    \begin{equation}\label{eq:derivative bound 1}
        z\big( (k-j)t \big) = \mathbf{R}_{jk}(t), \qquad \frac{d}{dt}z(t) = \frac{d}{dt} \mathbf{R}_{0k} \left(t/k\right).
    \end{equation}
    Using the expression \eqref{eq:relation between R and J} for $\frac{d}{dt}\mathbf{R}_{jk}(t)$, we can write the second equation in \eqref{eq:derivative bound 1} as
    \begin{equation}\label{eq:derivative bound 2}
        \begin{split}
            \frac{d}{dt}z(t) &= -\imath k \, \text{Tr} \big(\rho_0(t) \, [H,\rho_k(t)] \big) \\
            &= -\imath k \, \text{Tr} \big(\rho \, [H,\rho_k(t)] \big),
        \end{split}
    \end{equation}
    where the second line follows from the fact that $\rho_0(t) = U^0(t)\rho\,U^0(t) = \rho$. A simple algebra gives
    \begin{equation}\label{eq:derivative bound 3}
        \begin{split}
            \left|\frac{d}{dt}z(t)\right| = k \, \bigg| \text{Tr} \big([\rho,\rho_k(t)] \, H \big) \bigg|.
        \end{split}
    \end{equation}
    Define the $p$-Schatten norm of $A\in{\mathcal{M}_N}$ by
    \begin{equation}\label{eq:defn schatten norm}
    \begin{split}
        \|A\|_{S_p}& \coloneqq  \left( \sum_{i=0}^{N-1} \varsigma_i(A)^p \right)^{1/p} \quad \text{for} \quad p\in[1,\infty); \\
        \|A\|_{S_\infty}& \coloneqq  \varsigma_{N-1}(A),
    \end{split}
    \end{equation}
    where $\varsigma_i(A)$ denotes the $i$-th lowest singular value of $A$. The Schatten norm satisfies the following Hölder-type inequality: For all indices $p,q\in[0,\infty]$ with $1/p + 1/q = 1$,
    \begin{equation}\label{eq:schatten holder}
        \left| \text{Tr} (A^\dag B) \right| \leq  
        \|A\|_{S_p} \, \|B\|_{S_q}, \qquad \forall A,B \in {\mathcal{M}_N}.
    \end{equation}
    Applying the inequality \eqref{eq:schatten holder} to \eqref{eq:derivative bound 3}, we get
    \begin{equation}\label{eq:derivative bound 4}
        \begin{split}
            \left|\frac{d}{dt}z(t)\right| &\leq k \, \|[\rho,\rho_k(t)]\|_{S_1} \, \|H\|_{S_\infty}.
        \end{split}
    \end{equation}
    By the triangle inequality and sub-multiplicative property for $\|\cdot\|_{S_p}$, the first term in \eqref{eq:derivative bound 4} satisfies
    \begin{equation}\label{eq:derivative bound 5}
    \begin{split}
        \|[\rho,\rho_k(t)]\|_{S_1} &= \|\rho \,\rho_k(t) - \rho_k(t) \, \rho\|_{S_1} \\
        &\leq \|\rho \|_{S_1} \,\|\rho_k(t) \|_{S_1} + \| \rho_k(t) \|_{S_1} \, \| \rho\|_{S_1}.
    \end{split}
    \end{equation}
    Let $\lambda_i(\cdot)$ denote the $i$-th lowest eigenvalue of $(\cdot)$. Then $\varsigma_i(\rho_k(t)) = |\lambda_i(\rho_k(t))|$, since $\rho_k(t)$ are Hermitian. Furthermore, $\rho_k(t)$ are pure (rank-1) density matrices, so
    \begin{equation}\label{eq:derivative bound 6}
    \begin{split}
        \| \rho_k(t) \|_{S_1} &= \sum_{i=0}^{N-1}\varsigma_i(\rho_k(t)) \\
        &= \sum_{i=0}^{N-1}|\lambda_i(\rho_k(t))| \\
        &= 1.
    \end{split}
    \end{equation}
    Using \eqref{eq:derivative bound 6}, the upper bound in \eqref{eq:derivative bound 5} becomes $\|[\rho,\rho_k(t)]\|_{S_1} \leq 2$. In turn, \eqref{eq:derivative bound 4} is upper bounded as
    \begin{equation}\label{eq:derivative bound 8}
    \begin{split}
        \left|\frac{d}{dt}z(t)\right| &\leq 2 m \, \|H\|_{S_\infty},
    \end{split}
    \end{equation}
    where we have used $k\leq m$.

    Now suppose, $H=\sum_{j=1}^K \alpha_i H_i$, where $H_i$ are local Pauli operators and $\alpha_i\in\{w\in\mathbb{C}:\,|w|\leq 1\}$. Then
    \begin{align}
        \|H\|_{S_\infty} \leq \sum_{j=1}^K |\alpha_i| \|H_i\|_{S_\infty}
    \end{align}
    Since $\|H_i\|_{S_\infty}\leq 1$ and $|\alpha_i|\leq1$, then $\|H\|_{S_\infty}\leq K$. Plugging this upper bound on $\|H\|_{S_\infty}$ back into \eqref{eq:derivative bound 8} yields the desired result.
\end{proof}

\begin{example}
    Suppose $H$ consists of all nearest-neighbour interactions, i.e., $H=\sum_{i=0}^{n-1} \alpha_i H_{i,i+1}$. Then, by \cref{lemma:derivative upper bound},
    \begin{align}
        \left| \frac{d}{dt}z(t) \right|  \leq 2m (n-1).
    \end{align}
    Thus, the derivative of $z$ scales linearly in the number of qubits $n$ in the worst case.
\end{example}

\subsection{Proof of \cref{lemma:derivative-convergence}}\label{sec-app:derivative-estimate-convergence}

\noindent\textbf{Lemma \ref{lemma:derivative-convergence}.} \emph{Take $\widehat{x}$ as defined in \eqref{eq:min-problem-app}. In the limits $\|\eta\|_2\to0$ and $D\to\infty$, we have
    \begin{equation}\label{eq:x-convergence-lemma-app}
        \left\| \widehat{x} - x \right\|_\mathcal{V} \to 0.
    \end{equation}
}
\begin{proof}
    Write the minimization problem in \eqref{eq:min-problem-app} as
    \begin{equation}\label{eq:min-prob-obj-fn}
        \widehat{x}=\argmin_{  w \in \mathcal{D}(L)} Q(w), \qquad Q( {w})  \coloneqq  \| {y} - \widetilde{C} {w}\|^2_2 + \theta\|L {w}\|^2_\mathcal{V}.
    \end{equation}
    By the definition \eqref{eq:defn L} of $L$, the second term in $Q(w)$ can be expanded as
    \begin{equation}\label{eq:diff eqn for z}
        L {w}(t) = \frac{d}{dt}  {w}(t) - A  {w}(t), \qquad  {w}(0) =  {x}_{in},
    \end{equation}
    Define $ {f}_{ {w}}\in\mathcal{V}$ as the ``right-hand side'' of the operator equation \eqref{eq:diff eqn for z}, i.e., ${f}_{ {w}}  \coloneqq  L {w}$. 
    The differential equation \eqref{eq:diff eqn for z} is solved by
    \begin{equation}\label{eq:solution of equation for z(t)}
        {w}(t) = e^{At}  {x}_{in} + \int_\Omega e^{A(t-s)}  {f}_{ {w}}(s) ds.
    \end{equation} 
    Inserting the expression \eqref{eq:solution of equation for z(t)} into the argument of $Q$ yields
    \begin{equation}\label{eq:Mw 1}
        \begin{split}
            Q\big({w}\big) &= \left\| {y} - \widetilde{C}\left( e^{At}  {x}_{in} \right) + \widetilde{C}\left(\int_\Omega e^{A(t-s)}  {f}_{ {w}}(s) ds \right)\right\|^2_2 + \theta \big\|  {f}_{ {w}}\big\|^2_\mathcal{V},
        \end{split}
    \end{equation}
    where $Q(w)$ depends implicitly on $w$ through ${f}_{ {w}} = Lw$. We proceed to simplify \eqref{eq:Mw 1}. First, notice
    \begin{equation}\label{eq:Mw 2}
    \begin{split}
        \widetilde{C}\left(\int_\Omega e^{A(t-s)}  {f}_{ {w}}(s) ds \right)_{j} &= \left(\phi_j(t) {e}_0, \int_\Omega e^{A(t-s)} f_{ {w}}(s) ds\right)_{\mathcal{V}} \\
        &= \int_\Omega \chi_j(s)  {e}_0^{\,T} e^{A(t_j-s)}  f_{ {w}}(s) ds
    \end{split}
    \end{equation}
    where $\chi_j(t)$ is the Heaviside function
    \begin{equation}\label{eq:chi defn}
        \chi_j(t)  \coloneqq 
        \begin{cases}
            1 \quad \text{if} \,\, t \leq t_j \\
            0 \quad \text{if} \,\, t > t_j
        \end{cases}.
    \end{equation}
    Let ${k}^{(j)}(t)  \coloneqq  \chi_j(t) {e^{A^T(t_j-t)}}  {e}_0$. Then \eqref{eq:Mw 2} becomes
    \begin{equation}\label{eq:Mw 3}
    \begin{split}
        \widetilde{C}\left(\int_\Omega e^{A(t-s)}  {f}_{ {w}}(s) ds \right)_j  &= \int_\Omega  {k}^{(j)}(s)   {f}_{ {w}}(s) ds \\
        &= \left(  {k}^{(j)} ,  {f}_{ {w}}\right)_{\mathcal{V}}.
    \end{split}
    \end{equation}
    Inserting the expression \eqref{eq:Mw 3} back into \eqref{eq:Mw 1}, and using the definition of $\|\cdot\|_2$, we get
    \begin{equation}\label{eq:Mw 4}
        \begin{split}
            Q\big(w\big) &= \sum_{j=1}^{D} \left( {y}_j - \widetilde{C}\left( e^{At}  {x}_{in} \right)_j + \left(  {k}^{(j)} ,  {f}_{ {w}}\right)_{\mathcal{V}}\right)^2 + \theta \big\|  {f}_{ {w}}\big\|^2_\mathcal{V}.
        \end{split}
    \end{equation}
    Minimizing \eqref{eq:Mw 4} via direct differentiation, we get
    \begin{equation}\label{eq:Mw 5}
    \begin{split}  
        0 = - 2 \sum_{j=1}^{D} \left(\,  {y}_j - \widetilde{C}\left( e^{At}  {x}_{in} \right)_j - \big(  {k}^{(j)},  {f}_{\widehat{  x}}\big)_{\mathcal{V}} \, \right)  {k}^{(j)} + 2 \theta \,  {f}_{\widehat{  x}}
    \end{split}
    \end{equation}
    where ${f}_{\widehat{  x}} \coloneqq L \widehat{x}$, for $\widehat{x}$ solving \eqref{eq:min-prob-obj-fn}. Using the notation $\widetilde{ {y}}_j \coloneqq  {y}_j - \widetilde{C}\left( e^{At}  {x}_{in} \right)_j$ and rearranging \eqref{eq:Mw 5}, we get
    \begin{equation}\label{eq:Mw 6}
        \theta  {f}_{\widehat{  x}} = \sum_{j=1}^{D} \left(\, \widetilde{ {y}}_j - \big(  {k}^{(j)},  {f}_{\widehat{  x}}\big)_{\mathcal{V}} \, \right)  {k}^{(j)}.
    \end{equation}
    
    We now solve \eqref{eq:Mw 6} for $ {f}_{\widehat{  x}}$ to obtain an upper bound on the corresponding estimation error in $\widehat{x}$. Consider the ansatz
    \begin{equation}\label{eq:x-ansatz}
        {{f}}_{\widehat{x}} = \sum_{j=1}^{D} \beta_j  {k}^{(j)}
    \end{equation}
    for some $\beta_1,...,\beta_{D}\in\mathbb{R}$. Note that $k^{(j)}$ are linearly independent. Inserting \eqref{eq:x-ansatz} back into \eqref{eq:Mw 6} yields
    \begin{equation}\label{eq:Mw 7}
        \theta \sum_{j=1}^{D} \beta_j  {k}^{(j)} = \sum_{j=1}^{D} \sum_{i=1}^{D} \left(\, \widetilde{ {y}}_j - \beta_i\big(  {k}^{(j)},  {k}^{(i)}\big)_{\mathcal{V}} \, \right)  {k}^{(j)}.
    \end{equation}
    Equating the coefficients of $ {k}^{(j)}$ in \eqref{eq:Mw 7}, we have
    \begin{equation}
        \theta \beta_j  {k}^{(j)} = \widetilde{ {y}}_j - \sum_{i=1}^{D} \beta_i \big(  {k}^{(j)},  {k}^{(i)} \big)_{\mathcal{V}},
    \end{equation}
    or equivalently, in matrix notation, $\left( \theta I + \mathcal{K} \right)  {\beta} = \widetilde{ {y}}$, where $\mathcal{K}$ is the order-$D$ Gramian matrix $\mathcal{K}_{ij} \coloneqq \big( {k}^{(i)}, {k}^{(j)}\big)_{\mathcal{V}}$. Recall $\widetilde{ {y}}_j =  {y}_j - \widetilde{C}\left( e^{At}  {x}_{in} \right)_j$, so we may write the linear system in terms of the true function $x$:
    \begin{equation}\label{eq:beta linear sys 2}
    \begin{split}
        \left( \theta I + \mathcal{K} \right)  {\beta} &=  {y} - \widetilde{C}\left( e^{At}  {x}_{in} \right) \\
        &= \widetilde{C} {x}(t) +  \eta - \widetilde{C}\left( e^{At}  {x}_{in} \right) \\
        &= \widetilde{C}\left( {x}(t) - e^{At}  {x}_{in} \right) +  \eta.
    \end{split}
    \end{equation}
    Recall that $x$ satisfies the differential equation \eqref{eq:x-diff-app}, and so, similarly to \eqref{eq:solution of equation for z(t)}, 
    \begin{equation}\label{eq:beta linear sys 2a}
        {x}(t) = e^{At}  {x}_{in} + \int_\Omega e^{A(t-s)}  {f}(s) ds \quad \Rightarrow \quad {x}(t) - e^{At}  {x}_{in} = \int_\Omega e^{A(t-s)}  {f}(s) ds,
    \end{equation}
    where $f$ is the true ``right-hand side" corresponding to $x$, i.e., $f=Lx$. Using \eqref{eq:beta linear sys 2a}, then \eqref{eq:beta linear sys 2} becomes
    \begin{equation}\label{eq:beta linear sys 3}
    \begin{split}
        \left( \theta I + \mathcal{K} \right)  {\beta} &= \widetilde{C} \left( \int_\Omega e^{A(t-s)}  {f}(s) ds \right) +  \eta \\
        &= \bigg(\big( {k}^{(1)} ,  {f}\big)_{\mathcal{V}},...,\big(  {k}^{(D)} ,  {f}\big)_{\mathcal{V}}\bigg)^T +  \eta.
    \end{split}
    \end{equation}
    
    Take the limit $\|  \eta \,\|_2\to 0$. Furthermore, take $\theta_2\to\infty$ such that $\theta_2\|  \eta\|_2^2\to 0$ and fix $\theta_1>0$ such that \eqref{eq:theta-app-operator} is satisfied. Then $\theta = \theta_1/\theta_2 \to 0$ and \eqref{eq:beta linear sys 3} becomes
    \begin{equation}\label{eq:beta linear sys 4}
        \mathcal{K}  {\beta} = 
        \begin{pmatrix}
            \left(  {k}^{(1)} ,  {f}\right)_{\mathcal{V}} \\
            \vdots \\
            \left(  {k}^{(D)} ,  {f}\right)_{\mathcal{V}}
        \end{pmatrix}
        .
    \end{equation}
    It can easily verified that $\beta$ in \eqref{eq:beta linear sys 4} solves the minimization problem
    \begin{equation}\label{eq:beta zn min prob}
          \beta = \argmin_{\alpha\in\mathbb{R}^D} \bigg\| \sum_{j=1}^{D}\alpha_j   k^{(j)} -  {f}  \bigg\|_{\mathcal{V}}.
    \end{equation}
    Now take the limit $D\to\infty$. Then, since $  k^{(j)}$ are linearly independent, 
    \begin{equation}\label{eq:f convergence}
        \bigg\| \sum_{j=1}^{\infty}\beta_j k^{(j)} -  {f}  \bigg\|_{\mathcal{V}} \to 0 \qquad \Rightarrow \qquad \left\| {f}_{\widehat{x}} -   f  \right\|_\mathcal{V} = 
        \bigg\|\sum_{j=1}^{\infty} \beta_j   k^{(j)} -   f \, \bigg\|_\mathcal{V} \to 0.
    \end{equation}
    We have just to show that the corresponding estimate $\widehat{x}=L^{-1}f_{\widehat{x}}$ converges to the true function $x$. By \eqref{eq:solution of equation for z(t)},
    \begin{equation}\label{eq:full exp for hat x}
            \widehat{ {x}}(t) = e^{At}  {x}_{in} + \int_\Omega e^{A(t-s)}  {f}_{\widehat{ {x}}}(s) \,ds.
    \end{equation} 
    The estimation error can then be written
    \begin{equation}\label{eq:x hat convergence bound}
    \begin{split}
        \big\|  x - \widehat{ {x}}\big\|_\mathcal{V} &= \left\| e^{At}  {x}_{in} + \int_\Omega e^{A(t-s)}  {f}_{  x}(s) \,ds -  e^{At}  {x}_{in} - \int_\Omega e^{A(t-s)}  {f}_{\widehat{ {x}}}(s) \,ds \right\|_\mathcal{V} \\
        &= \left\| \int_\Omega e^{A(t-s)} \big( {f}(s) -  {f}_{\widehat{ {x}}}(s)\big) ds \right\|_\mathcal{V}.
    \end{split}
    \end{equation}
    Jensen's inequality states that, for a measurable function $p:\Omega\to\mathbb{R}$ and convex function $q:\mathbb{R} \to \mathbb{R}$,
    \begin{equation}
        q\left(\int_\Omega p(s) ds\right) \leq \int_\Omega q\big(p(s)\big) ds.
    \end{equation}
    Applying Jensen's inequality to \eqref{eq:x hat convergence bound}, for $q=\|\cdot\|_\mathcal{V}$ and $p(s) = e^{A(t-s)}( {f}(s) -  {f}_{\widehat{ {x}}}(s))$ for some fixed $t\in\mathbb{R}$, we get
    \begin{equation}
    \begin{split}\label{eq:x hat convergence bound 1}
        \big\|  x - \widehat{ {x}}\big\|_\mathcal{V} &\leq \int_\Omega \left\|  e^{A(t-s)} \big( {f}(s) -  {f}_{\widehat{ {x}}}(s)\big) \right\|_\mathcal{V} \,\, ds \\
        &\leq \int_\Omega \left\|  e^{A(t-s)} \right\|_\mathcal{V} \,\, \left\|  {f} -  {f}_{\widehat{ {x}}} \right\|_\mathcal{V} ds .
    \end{split}
    \end{equation}
    Finally, by \eqref{eq:f convergence}, the right-hand side of \eqref{eq:x hat convergence bound 1} tends to zero, giving the desired result in \eqref{eq:x-convergence-lemma-app}.
\end{proof}

\subsection{Proof of \cref{thm:full minimax expression}}\label{subsec-app:minimax-thm-proof}

\noindent \textbf{Theorem \ref{thm:full minimax expression}.} \emph{
Let $\phi_s(t)=\delta(t-t_s)$ be the generalized Dirac delta function centered at the data point $t_s$. Let $e_j$ be the vector with $j$-th entry equal to $1$ and all other entries equal to $0$. The solution $\widehat{{x}}$ of the minimax problem \eqref{eq:general minimax problem} solves
\begin{equation}\label{eq:minimax-sol-ODE-app}
\begin{split}
    \frac{d}{dt} \widehat{x}(t) &= A \widehat{x}(t) + \frac{1}{\theta_1} e_{M-1} e_{M-1}^T b(t), \\
    \frac{d}{dt} b(t) &= - A^T b(t) - \theta_2 \sum_{s=1}^{D} \phi_s(t) e_0 \left( y_s - (\phi_s e_0, \widehat{x})_\mathcal{V} \right),
\end{split}
\end{equation}
for some $b\in\mathcal{V}$, with boundary conditions $\widehat{{x}}(0)=x_{in}$ and ${b}(T) = 0$. Furthermore, for $l\in\mathcal{V}$, the corresponding minimax error \eqref{eq:general minimax error} in $\widehat{x}$ is
\begin{equation}
    \Upsilon_{\widehat{ x}}(l) = \sqrt{\left(l, \widehat{p}\right)_\mathcal{V}}
\end{equation}
where $\widehat{p}\in\mathcal{V}$ solves
\begin{equation}\label{eq:minimax-err-ODE-app}
\begin{split}
    \frac{d}{dt} \widehat{p}(t) &= A \widehat{p}(t) + \frac{1}{q} {e}_{M-1} e_{M-1}^T g(t), \\
    \frac{d}{dt} g(t) &= -A^T g(t) - l(t) + \theta_2 \sum_{s=1}^{D} ( \phi_s e_0, \widehat{p})_\mathcal{V} \, \phi_s(t)\,  e_0,
\end{split}
\end{equation}
for some $g\in\mathcal{V}$, with boundary conditions $\widehat{p}(0) = 0$ and $g(T) = 0$.
}

Before proceeding, we introduce some notation. Recall from \eqref{eq:x-diff-app} that the true function $x$ solves the differential equation
\begin{equation}\label{eq:full x eqn}
    \frac{d}{dt} x(t) = A  x(t) +  f(t).
\end{equation}
Now, define $\widetilde{x},\overline{x}\in\mathcal{V}$ solving the respective differential equations
\begin{equation}\label{eq:tilde x linear system}
    \frac{d}{dt} \widetilde{ x}(t) = A \widetilde{ x}(t), \quad \widetilde{ x}(0) =  x_{in},
\end{equation}
and
\begin{equation}\label{eq:bar x linear system}
    \frac{d}{dt} \overline{ x}(t) = A \overline{ x}(t) +  f(t), \quad \overline{ x}(0) =  0.
\end{equation}
Define the adjusted measurements
\begin{equation}\label{eq:shifted-msmts}
    \overline{y}  \coloneqq  \widetilde{C}\overline{ x} +  \eta \qquad  \textup{and} \qquad \widetilde{y}  \coloneqq  \widetilde{C}\widetilde{ x} + \eta
\end{equation}
Then, the true function $x$ and corresponding measurements $y$ can be written
\begin{equation}\label{eq:x decomp}
    x = \widetilde{x} + \overline{x} \qquad  \textup{and} \qquad y =  \widetilde{y} + \overline{y}.
\end{equation}

The proof of \cref{thm:full minimax expression} is organized as follows. We first construct a minimax estimate $\widehat{ \varphi}$ of the state $\overline{ x}$ in \eqref{eq:bar x linear system}, which call we call the \textit{homogeneous minimax estimate}, since $\overline{x}(0) =  0$. In \cref{lemma:u_hat minimax solution,lemma:minimax error of reformulated problem}, we show that the estimate $\widehat{ \varphi}$ and the corresponding minimax error, $\Upsilon_{\widehat{\varphi}}(l) \coloneqq \max_{w\in\mathcal{G}}|(l,\widehat{\varphi} - w)_{\mathcal{V}
}|$, can each be expressed as the solution of a first-order coupled differential equation. In \cref{thm:minimax error}, we show that the minimax error $\Upsilon_{\widehat{x}}$ of the original minimax problem in \eqref{eq:general minimax error} is equal to that of the homogeneous problem. Furthermore, we use the decomposition \eqref{eq:x decomp} to obtain an expression for the original minimax estimate $\widehat{x}$ in \eqref{eq:general minimax problem}.

The homogeneous minimax problem for $\overline{x}$ is defined as follows. Recall the definition of $\mathcal{G}_y$ from \eqref{eq:defn bounding space}: 
\begin{equation}
     \mathcal{G}_y \coloneqq \bigg\{ w \in \mathcal{V}:\,\, \theta_2 \,\|y - \widetilde{C} w\|^2_{2} + \theta_1 \, \|Lw\|^2_{\mathcal{V}} \leq 1 \bigg\}.
\end{equation}
We call $\mathcal{G}_y$ the \textit{space of admissible solutions} corresponding to the measurements $y$ (of $x$). Consider the admissible solutions relating to the measurements $\overline{y}$ (of $\overline{x}$):
\begin{equation}
    \mathcal{G}_{\overline{y}} \coloneqq \bigg\{ w \in \mathcal{V}:\,\, \theta_2 \,\|\overline{y} - \widetilde{C} w\|^2_{2} + \theta_1 \, \|L {w}\|^2_{\mathcal{V}} \leq 1 \bigg\}
\end{equation}
Then, for $l \in \mathcal{V}$, define the homogeneous minimax estimate $\widehat{\varphi} \in \mathcal{G}_{\overline{y}}$ by
\begin{equation}\label{eq:minimax prob for bar x}
    \max_{w \in \mathcal{G}_{\overline{y}}} \left| ( l, \widehat{\varphi} -  w)_{\mathcal{V}} \right| =
    \min_{q \in \mathcal{G}_{\overline{y}}} \max_{w \in \mathcal{G}_{\overline{y}}} \left| ( l, q - w)_{\mathcal{V}} \right|,
\end{equation}
with associated minimax error
\begin{equation}\label{eq:minimax error for bar x}
    \Upsilon_{\widehat{\varphi}}(l) = \max_{w \in \mathcal{G}_{\overline{y}}} \left| ( l, \widehat{\varphi} -  w)_{\mathcal{V}} \right|.
\end{equation}
In order to derive an expression for $\widehat{\varphi}$, we must first consider the following related problem: find a set of scalars $\widehat{u}_1,...,\widehat{u}_{D}\in\mathbb{R}$ such that, for all $l \in \mathcal{V}$,
\begin{equation}\label{eq:minimax for u_hat}
     \max_{w \in \mathcal{G}_{\overline{y}}} \left| \sum_{s=1}^{D} \widehat{u}_s \overline{y}_s -  (l, w)_{\mathcal{V}}  \right|
     =
     \min_{u_1,...,u_{D} \in \mathbb{R}} \max_{w \in \mathcal{G}_{\overline{y}}} \left| \sum_{s=1}^{D} u_s \overline{y}_s - (l,w)_{\mathcal{V}} \right|.
\end{equation}
with associated minimax error
\begin{equation}\label{eq:u minimax error}
    \max_{w \in \mathcal{G}_{\overline{y}}} \left| \sum_{s=1}^{D} \widehat{u}_s \overline{y}_s -  ( l, w)_{\mathcal{V}}  \right|.
\end{equation}

\cref{lemma:u_hat minimax solution} shows that the solution $\widehat{u} = (\widehat{u}_1,...,\widehat{u}_{D})^T$ of \eqref{eq:minimax for u_hat} can be expressed in terms of a coupled first-order differential equation. In \cref{lemma:minimax error of reformulated problem}, we show that, for any $l\in\mathcal{V}$, the corresponding minimax error \eqref{eq:u minimax error} is exactly equal to the minimax error \eqref{eq:minimax error for bar x} of $\widehat{\varphi}$.

\begin{lemma}\label{lemma:u_hat minimax solution}
    Take $\theta_1,\theta_2$ as defined in \eqref{eq:theta-app-operator}. The components of the minimax solution $\widehat{u}$ to \eqref{eq:minimax for u_hat} satisfy
    \begin{equation}\label{eq:u hat s lemma}
        \widehat{u}_s = \theta_2 (\phi_s e_0, \widehat{p}\,)_\mathcal{V},
    \end{equation}
    for $s=1,...,D$, where $\widehat{p}\in\mathcal{V}$ solves
    \begin{equation}\label{eq:system for z_hat apnx}
    \begin{split}
        \frac{d}{dt} \widehat{p}(t) &= A \widehat{p}(t) + \frac{1}{\theta_1} e_{M-1} e_{M-1}^T g(t), \\
        \frac{d}{dt} g(t) &= -A^T g(t) - l(t) + \theta_2 \sum_{s=1}^{D} \phi_s(t) e_0  ( \phi_s e_0, \widehat{p} \,)_\mathcal{V}\, ,
    \end{split}
    \end{equation}
    for some $g\in\mathcal{V}$, with boundary conditions $\widehat{p}(0) = 0$ and $g(T) = 0$.
    \begin{proof}
        Inserting the definition \eqref{eq:shifted-msmts} of $\overline{y}_s$ into the argument on the right-hand side of \eqref{eq:minimax for u_hat}, we get
        \begin{equation}\label{eq:lTx}
        \begin{split}
            \int_\Omega l^T(t) \overline{x}(t) dt &- \sum_{s=1}^{D} u_s \overline{y}_s = \int_\Omega l^T(t) \overline{x}(t) dt - \int_\Omega \sum_{s=1}^{D} \phi_s(t) e_0^T \overline{x}(t) \widehat{u}_s dt - \sum_{s=1}^{D} u_s \eta_s.
        \end{split}
        \end{equation}
        Take $g\in\mathcal{V}$ satisfying
        \begin{equation}
            \frac{d}{dt} g(t) = -A^T g (t) - l(t) + \sum_{s=1}^{D} \phi_s(t) e_0 u_s, \quad g(T) = 0.
        \end{equation}
        Notice
        \begin{equation}\label{eq:g tau x tau}
            0 = g^T(T) \overline{x}(T) = g^T(0) \overline{x}(0) + \int_\Omega \left\{ \left(\frac{d}{dt} g(t)\right)^T \, \overline{x}(t) + g^T(t) \left(\frac{d}{dt} \overline{x}(t)\right) \right\} dt.
        \end{equation}
        for $\overline{x}$ as defined in \eqref{eq:bar x linear system}. Then, adding $g^T(T)  \, \overline{x}(T)$ to the right-hand side of \eqref{eq:lTx} gives
        \begin{equation}\label{eq:lhs minmax x bar}
        \begin{split}
            \int_\Omega l^T(t) \overline{x}(t) dt &- \sum_{s=1}^{D} u_s \overline{y}_s = g^T(0) \,\overline{x}(0) + \int_\Omega \left(-A^T g(t) - l(t) + \sum_{s=1}^{D} \phi_s(t) e_0 u_s \right)^T \overline{x}(t) dt \\
            &\quad\quad + \int_\Omega g^T(t) \,(A\overline{x}(t) + f(t)) dt - \int_\Omega \sum_{s=1}^{D} \phi_s(t) e_0^T \overline{x}(t) u_s \, dt + \int_\Omega l^T(t) \overline{x}(t) dt - \sum_{s=1}^{D} u_s \eta_s \\
            &= \int_\Omega g^T(t) f(t) dt - \sum_{s=1}^{D} u_s \eta_s.
        \end{split}
        \end{equation}
        Using the Cauchy-Schwarz inequality and the inequality \eqref{eq:theta-app-operator} for $f$ and $\eta=y-\widetilde{C}x$, we can upper bound \eqref{eq:lhs minmax x bar} by 
        \begin{equation}
             \int_\Omega l^T(t) \overline{x}(t) dt - \sum_{s=1}^{D} u_s \overline{y}_s \leq \left( \underbrace{ \int_\Omega \left( \frac{1}{\theta_1} e_{M-1}^T g(t) \right)^T \left(e_{M-1}^T g(t)\right) dt + \frac{1}{\theta_2} \sum_{s=1}^{D} u_s^2}_{\scalebox{1.1}{~\hspace{0.25in}$=:\mathcal{J}(u_1,...,u_{D})$}} \right)^{1/2}.
        \end{equation}
        Thus, the minimax problem in \eqref{eq:minimax for u_hat} can be written equivalently as
        \begin{equation}\label{eq:minimax-equiv}
            \min_{u_1,...,u_{D} \in\mathbb{R}} \left\{ \max_{f,\eta \in \mathcal{E}} \left| \int_\Omega l^T(t) \overline{x}(t) dt - \sum_{s=1}^{D} u_s \overline{y}_s \right| \right\} = \min_{u_1,...,u_{D}\in\mathbb{R}} \mathcal{J}(u_1,...,u_{D}).
        \end{equation}
        \cref{eq:minimax-equiv} implies that the solution $\widehat{u}$ of \eqref{eq:minimax for u_hat} satisfies
        \begin{equation}\label{eq:defn uhat}
            {\widehat{u} \coloneqq \argmin_{u\in\mathbb{R}^D}\mathcal{J}(u)}.
        \end{equation}
        
         We proceed to solve the minimization problem \eqref{eq:minimax-equiv} analytically in order to obtain the desired expression for $\widehat{u}$. The directional derivative of $\mathcal{J}$ reads
        \begin{equation}\label{eq: ddJu}
            \frac{d}{d\gamma} \mathcal{J}(u + \gamma v) = 2 \left\{ \int_\Omega \left( \frac{1}{\theta_1} e_{M-1}^T r'(t) \right)^T e_{M-1}^T r_\gamma(t) dt +  \frac{1}{\theta_2} \sum_{s=1}^{D} (u_s + \gamma v_s) v_s \right\},
        \end{equation}
        for $\gamma\in\mathbb{R}$, $v\in\mathbb{R}^D$, where $r_\gamma \in \mathcal{V}$ solves
        \begin{align}\label{eq:z prime diff}
            \frac{d}{dt} r_\gamma(t) = -A^T r_\gamma (t) + l(t) + \sum_{s=1}^{D} \phi_s(t) e_0 (u_s + \gamma v_s), \qquad\qquad r_\gamma(T) = 0,
        \end{align}
        and $r' \in \mathcal{V}$ solves
        \begin{equation}
            \frac{d}{dt} r'(t) = -A^T r' (t) + \sum_{s=1}^{D} \phi_s(t) e_0 v_s, \qquad\qquad r'(T) = 0.
        \end{equation}
        Now, set $\gamma=0$. Then $g(t) = r_{\gamma=0}(t)$ and \eqref{eq: ddJu} becomes
        \begin{equation}\label{eq:dJdgamma eval}
            \frac{d}{d\gamma} \mathcal{J}(u + \gamma v) \biggr|_{\gamma=0} = 2 \left\{ \int_\Omega \left(\frac{1}{\theta_1} e_{M-1}^T r'(t) \right)^T e_{M-1}^T g(t) dt + \frac{1}{\theta_2} \sum_{s=1}^{D} u_s v_s \right\}.
        \end{equation}
        Let $\widehat{p}\in\mathcal{V}$ satisfy \eqref{eq:system for z_hat apnx}. Then \eqref{eq:dJdgamma eval} can be written as
        \begin{equation}\label{eq:d J d gamma calc}
            \frac{d}{d\gamma} \mathcal{J}(\widehat{u} + \gamma v) \biggr|_{\gamma=0} = 2 \left\{ r'\,^T(0) \, \widehat{p}(0) + \int_\Omega \left( \frac{d}{dt} \widehat{p}(t) - A \widehat{p}(t) \right)^T r'(t) dt + \frac{1}{\theta_2} \sum_{s=1}^{D} u_s v_s \right\} 
        \end{equation}
        Inserting the respective expressions \eqref{eq:system for z_hat apnx} and \eqref{eq:z prime diff} for $\frac{d}{dt}\widehat{p}(t)$ and $\frac{d}{dt}r'(t)$ into \eqref{eq:d J d gamma calc}, we get
        \begin{equation}\label{eq:d dt J}
        \begin{split}
        \frac{d}{d\gamma} \mathcal{J}(\widehat{u} + \gamma v) \biggr|_{\gamma=0} & = 2 \left\{ r'\,^T(T) \widehat{p}(T) -  \int_\Omega \widehat{p}^T(t) \left( {-A^T r'(t)} + \sum_{s=1}^{D} \phi_s(t) e_0 v_s \right) dt - \int_\Omega {(A \widehat{p}(t))^T r'(t)} dt + \frac{1}{\theta_2} \sum_{s=1}^{D} u_s v_s \right\} \\
            &= 2 \left\{ \sum_{s=1}^{D} \int_\Omega \phi_s(t) e_0^T \widehat{p}(t) dt \, v_s + \frac{1}{\theta_2} \sum_{s=1}^{D} u_s v_s \right\}. 
        \end{split} 
        \end{equation}
        The expression for $\widehat{u}$ in \eqref{eq:u hat s lemma} is obtained by setting \eqref{eq:d dt J} equal to zero and equating the coefficients of $v_s$.
    \end{proof}
\end{lemma}

\begin{lemma}\label{lemma:minimax error of reformulated problem}
    The minimax solution $\,\sum_{s=1}^{D} \widehat{u}_s \overline{y}_s$ to \eqref{eq:minimax for u_hat} satisfies
    \begin{equation}\label{eq:u hat lemma}
        \sum_{s=1}^{D} \widehat{u}_s \overline{y}_s = ( l, \widehat{\varphi})_\mathcal{V}
    \end{equation}
    where $\widehat{\varphi}\in\mathcal{V}$ solves
    \begin{equation}\label{eq:phi hat and w}    
    \begin{split}
        \frac{d}{dt} \widehat{\varphi}(t) &= A \widehat{\varphi}(t) + \frac{1}{\theta_1} e_{M-1} e_{M-1}^T w(t), \\
        \frac{d}{dt} w(t) &= - A^T w(t) - \theta_2 \sum_{s=1}^{D} \phi_s(t) e_0 \bigg( \overline{y}_s - ( \phi_s e_0, \widehat{\varphi})_\mathcal{V} \bigg),
    \end{split}
    \end{equation}
    for some $w\in\mathcal{V}$, with boundary conditions $\widehat{\varphi}(0)=0$ and $w(T) = 0$. Furthermore, the minimax error in \eqref{eq:u minimax error} can be written as
    \begin{equation}
        \max_{w \in \mathcal{G}_{\overline{y}}} \left| \sum_{s=1}^{D} \widehat{u}_s \overline{y}_s -  ( l, w)_{\mathcal{V}}  \right| = \max_{w \in \mathcal{G}_{\overline{y}}} \left|
        (l,\widehat{\varphi} - w)_\mathcal{V} \right|
        = \sqrt{( l, \widehat{p})_\mathcal{V}},
    \end{equation}
    where $\widehat{p}\in\mathcal{V}$ solves \eqref{eq:system for z_hat apnx}.
    \begin{proof}
        We prove the result by direct calculation. Using the expression \eqref{eq:u hat s lemma} for $\widehat{u}_s$, then \eqref{eq:u hat lemma} can be written
        \begin{equation}\label{eq:sub 1}
            \sum_{s=1}^{D} \widehat{u}_s \overline{y}_s = \int_\Omega \widehat{p}^T(t) \left( \theta_2 \sum_{s=1}^{D} \phi_s(t) e_0 \overline{y}_s \right) dt
        \end{equation}
        Take $(\widehat{p},g)$ satisfying \eqref{eq:system for z_hat apnx} and $(\widehat{\varphi},w)$ satisfying \eqref{eq:phi hat and w}. Then, by the expression \eqref{eq:phi hat and w} for $w$, we can write \eqref{eq:sub 1} as
        \begin{equation}\label{eq:sub 2}
        \begin{split}
            \sum_{s=1}^{D} \widehat{u}_s \overline{y}_s &= \int_\Omega \widehat{p}^T(t) \left( - \frac{d}{dt} w(t) - A^T w(t) + \theta_2 \sum_{s=1}^{D} \phi_s(t) e_0 \int_\Omega \phi_s(t') e_0^T \widehat{\varphi}(t') dt' \right) dt \\ 
            &= - \int_\Omega \widehat{p}^T(t) \left(\frac{d}{dt} w(t)\right) dt - \int_\Omega \widehat{p}^T(t) A^T w(t) dt \\
            & \qquad\qquad + \theta_2\sum_{s=1}^{D} \left(\int_\Omega \phi_s(t) e_0^T \widehat{\varphi}(t) dt \int_\Omega \phi_s(t') e_0^T \widehat{p}(t') dt'\right).
        \end{split}
        \end{equation}
        We insert the expression \eqref{eq:system for z_hat apnx} for $\widehat{p}$ into \eqref{eq:sub 2} and rearrange to get
        \begin{equation}
        \label{eq:sub 3}
        \begin{split}
            \sum_{s=1}^{D} \widehat{u}_s \overline{y}_s &= - \left( \widehat{p}^T(T) w(T) - \widehat{p}^T(0) w(0) \right)  + \int_\Omega \left( {A \widehat{p}(t)} + \frac{1}{\theta_1} e_{M-1} e_{M-1}^T g(t) \right)^T w(t) dt \\
            & \quad\quad- \int_\Omega {\widehat{p}^T(t) A^T w(t)} dt + \int_\Omega \widehat{\varphi}^T(t) e_0 \sum_{s=1}^{D} \theta_2 \left( \int_\Omega \phi_s(t') e_0^T \widehat{p}(t') dt'\right) \phi_s(t) dt.
        \end{split}
        \end{equation}
        Since $g$ solves \eqref{eq:system for z_hat apnx}, we can write \eqref{eq:sub 3} as
        \begin{equation}
        \label{eq:sub 4}
            \sum_{s=1}^{D} \widehat{u}_s \overline{y}_s = \int_\Omega g^T(t) \left( \frac{d}{dt} \widehat{\varphi}(t) - {A \widehat{\varphi}(t)} \right) dt + \int_\Omega \widehat{\varphi}^T(t) \left( \frac{d}{dt} g(t) + {A^T g(t)} + l(t) \right) dt.
        \end{equation}
        Notice that
        \begin{equation}\label{eq:sub 4a}
            \widehat{\varphi}^T(T) g(T) - \widehat{\varphi}^T(0) g(0) = \int_\Omega \left\{ \left(\frac{d}{dt} \widehat{\varphi}(t)\right)^T \, g(t) + \widehat{\varphi}^T(t) \left(\frac{d}{dt} g(t)\right) \right\} dt.
        \end{equation}
        Inserting the expression \eqref{eq:sub 4a} back into \eqref{eq:sub 4} gives
        \begin{equation}\label{eq:sub 5}
            \sum_{s=1}^{D} \widehat{u}_s \overline{y}_s = \widehat{\varphi}^T(T) g(T) - \widehat{\varphi}^T(0) g(0) + \int_\Omega \widehat{\varphi}^T(t) l(t)dt.
        \end{equation}
        Since $g(T)=0$, $\widehat{\varphi}(0)=0$, then \eqref{eq:sub 5} simplifies to
        \begin{equation}
        \label{eq:sub 6}
            \sum_{s=1}^{D} \widehat{u}_s \overline{y}_s = \int_\Omega \widehat{\varphi}^T(t) l(t)dt.
        \end{equation}
        Recall from \cref{lemma:u_hat minimax solution} that
        \begin{equation}\label{eq:sub here}
        \min_{u_1,...,u_{D}\in\mathbb{R}} \, \max_{f,\eta \in \mathcal{E}} \left| \int_\Omega l^T(t') \overline{x}(t') dt' - \sum_{s=1}^{D} u_s \overline{y}_s \right| = \min_{u_1,...,u_{D}\in\mathbb{R}} \mathcal{J}(u_1,...,u_{D}) = \mathcal{J}(\widehat{u}_1,...,\widehat{u}_{D})
        \end{equation}
        where $\widehat{u}_s = \theta_2 (\phi_s e_0, \widehat{p} \,)_\mathcal{V}$. Then, inserting the expression \eqref{eq:sub 6} into the argument of the minimax problem \eqref{eq:sub here} yields
        \begin{equation}\label{eq:sub 1a}
            \int_\Omega l^T(t)\widehat{p}(t)dt = \mathcal{J}(\widehat{u}_1,...,\widehat{u}_{D}).
        \end{equation}
        Finally, plugging \eqref{eq:sub 6}, \eqref{eq:sub 1a} back into \eqref{eq:sub here} yields the desired result:
        \begin{equation}
            \min_{u_1,...,u_{D} \in\mathbb{R}} \, \max_{f,\eta \in \mathcal{E}} \left| \int_\Omega l^T(t) \, \overline{x}(t) dt - \int_\Omega l^T(t) \, \widehat{\varphi}(t) dt \right| = \int_\Omega l^T(t) \, \widehat{p}(t)dt.
        \end{equation}
    \end{proof}
\end{lemma}

The following Lemma connects the minimax estimate $\widehat{x}$ and error $\Upsilon_{\widehat{x}}$ from the original problem \eqref{eq:general minimax problem} to the minimax estimate $\widehat{\varphi}$ and error $\Upsilon_{\widehat{\varphi}}$ in the problem \eqref{eq:minimax prob for bar x}. This allows us to express each of $\widehat{x}$ and $\Upsilon_{\widehat{x}}$ as the solution of a coupled first-order differential equation, thus providing a means of analytically computing each of these quantities.

\begin{lemma}\label{thm:minimax error}
    Take $\widehat{x}$ and $\Upsilon_{\widehat{x}}$ as defined in \eqref{eq:general minimax problem} and \eqref{eq:general minimax error} respectively. The estimate $\widehat{x}$ solves
    \begin{equation}\label{eq:minimax-err-diff-eq}
    \begin{split}
        \frac{d}{dt} \widehat{x}(t) &= A \widehat{x}(t) + \frac{1}{\theta_1} e_{M-1} e_{M-1}^T b(t), \\
        \frac{d}{dt} b(t) &= - A^T b(t) - \theta_2 \sum_{s=1}^{D} \phi_s(t) e_0 \bigg( y_s - (\phi_s e_0, \widehat{x})_\mathcal{V} \bigg),
    \end{split}
    \end{equation}
    for some $b\in\mathcal{V}$, with boundary conditions $\widehat{x}(0)=x_{in}$ and $b(T) = 0$. Let $\widehat{p}$ solve \eqref{eq:system for z_hat apnx}. Then the minimax error $\Upsilon_{\widehat{x}}(l)$ can be written
    \begin{equation}
        \Upsilon_{\widehat{x}}(l)  \coloneqq  \max_{w \in \mathcal{G}_{\overline{y}}} \left|
        (l,\widehat{x} - w)_\mathcal{V}\right| = \sqrt{\left(l, \widehat{p}\,\right)_\mathcal{V}}.
    \end{equation}
    \end{lemma}    
    \begin{proof}
        Recall from \eqref{eq:x decomp} that $x =\overline{x} + \widetilde{x}$. By the definition \eqref{eq:shifted-msmts} of $\overline{y}$, we have $y = \overline{y} + \widetilde{C}\widetilde{x}$. Thus, we can write the minimax estimate $\widehat{x}$ of $x$ in \eqref{eq:general minimax problem} as $\widehat{x} = \widehat{\varphi} + \widetilde{x}$, where $\widehat{\varphi}$ is ``homogeneous'' minimax estimate of $\overline{x}$ in \eqref{eq:minimax prob for bar x}. Furthermore, we have
        \begin{equation}\label{eq:ddt x hat}
            \frac{d}{dt}\widehat{x}(t) = \frac{d}{dt} \widehat{\varphi} (t) + \frac{d}{dt}\widetilde{x}(t).
        \end{equation}
        Recall from \eqref{eq:tilde x linear system} that $\widetilde{x}$ solves
        \begin{equation}\label{eq:ddt tilde x}
            \frac{d}{dt} \widetilde{x}(t) = A \widetilde{x}(t), \qquad \widetilde{x}(0) = x_{in}.
        \end{equation} 
        Moreover, recall from \cref{lemma:minimax error of reformulated problem} that the pair $(\widehat{\varphi},w)\in\mathcal{V}\times\mathcal{V}$ solves
        \begin{equation}\label{eq:varphi diff eq}
        \begin{split}
        \frac{d}{dt} \widehat{\varphi}(t) &= A \widehat{\varphi}(t) + \frac{1}{\theta_1} e_{M-1} e_{M-1}^T w(t), \\
        \frac{d}{dt} w(t) &= - A^T w(t) - \theta_2 \sum_{s=1}^{D} \phi_s(t) e_0 \left( \overline{y}_s - ( \phi_s e_0, \widehat{\varphi})_\mathcal{V} \right),
        \end{split}
        \end{equation}
        for $\widehat{\varphi}(0)=0$ and $w(T) = 0$. Using $\widehat{\varphi} = \widehat{x} - \widetilde{x}$, the second equation in \eqref{eq:varphi diff eq} can be written
        \begin{equation}\label{eq:ddt w}
            \frac{d}{dt} w(t) = - A^T w(t) - \theta_2 \sum_{s=1}^{D} \phi_s(t) e_0 \bigg( \overline{y}_s - ( \phi_s e_0, \widehat{x})_\mathcal{V} + ( \phi_s e_0,\widetilde{x})_\mathcal{V} \bigg).
        \end{equation}
        From the definition \eqref{eq:defn-Ctilde} of $\widetilde{C}$, we have $( \phi_s e_0,\widetilde{x})_\mathcal{V} = \widetilde{C}(\widetilde{x})_s$, the $s$-th component of the vector $\widetilde{C}(\widetilde{x})$. Furthermore, by \eqref{eq:x decomp} we have $\widetilde{C}(\widetilde{x}) = y - \overline{y}$, and so \eqref{eq:ddt w} becomes
        \begin{equation}
        \begin{split}
            \frac{d}{dt} w(t) &= - A^T w(t) - \theta_2 \sum_{s=1}^{D} \phi_s(t) e_0 \bigg( \overline{y}_s - ( \phi_s e_0, \widehat{x})_\mathcal{V} + y_s - \overline{y}_s \bigg) \\
            &= A^T w(t) - \theta_2 \sum_{s=1}^{D} \phi_s(t) e_0 \bigg( y_s - ( \phi_s e_0, \widehat{x})_\mathcal{V} \bigg).
        \end{split}
        \end{equation}
        In order to obtain the second equation in \eqref{eq:minimax-err-diff-eq}, we insert \eqref{eq:ddt tilde x} and the second equation in \eqref{eq:varphi diff eq} back into the expression \eqref{eq:ddt x hat}, yielding
        \begin{equation}
        \begin{split}
             \frac{d}{dt}\widehat{x}(t) &= A \widehat{\varphi}(t) + \frac{1}{\theta_1} e_{M-1} e_{M-1}^T w(t) + A\widetilde{x}(t)\\
             &= A \widehat{x}(t) + \frac{1}{\theta_1} e_{M-1} e_{M-1}^T w(t).
        \end{split}
        \end{equation}

        Finally, for the minimax error $\Upsilon_{\widehat{x}}$, insert $\widehat{\varphi} = \widehat{x} - \widetilde{x}$ and $\overline{x} = x - \widetilde{x}$ into the definition of $\Upsilon_{\widehat{\varphi}}$ in \eqref{eq:minimax error for bar x} to get
        \begin{equation}\label{eq:final sigma hat var}
        \begin{split}
            \Upsilon_{\widehat{\varphi}}(l) & =  \max_{\overline{x}\in\mathcal{G}_{\overline{y}}} \left| (l,\widehat{\varphi})_\mathcal{V} - (l,\overline{x})_\mathcal{V} \right| \\
            &=\max_{x\in\mathcal{G}_{\overline{y}}} \left| (l,\widehat{x} + \widetilde{x})_\mathcal{V} - (l,x - \widetilde{x})_\mathcal{V} \right| \\
            &= \max_{x \in \mathcal{G}_{\overline{y}}} \left| (l, \widehat{x} - x\,)_{\mathcal{V}} \right| \\
            &= \Upsilon_{\widehat{x}}(l),
        \end{split}
        \end{equation}
        for $l\in\mathcal{V}$, where in the last line we have used the definition \eqref{eq:general minimax error} of $\Upsilon_{\widehat{x}}$. From \cref{lemma:minimax error of reformulated problem} we have $\Upsilon_{\widehat{\varphi}}(l)=\sqrt{\left(l, \widehat{p}\,\right)_\mathcal{V}}$, which completes the proof.
    \end{proof}

\vspace{2cm}

\bibliographystyle{unsrt}
\bibliography{references}

\end{document}